\documentclass[twocolumn,trackchanges]{aastex63}
\RequirePackage{lineno}
\usepackage{amsmath}
\usepackage{amssymb}
\usepackage{amsthm}
\usepackage{natbib,aasdefs,url,bm}
\usepackage{array}
\usepackage{float}
\usepackage{graphicx}
\usepackage{subfigure}
\usepackage{color}
\usepackage{threeparttable} % to use table notes
\usepackage{CJK}
\usepackage{lineno}
\usepackage{tablefootnote}
\usepackage{threeparttable}

\newcommand{\equ}[1]{Eq.~(\ref{equ:#1})}
\newcommand{\figu}[1]{Fig.~\ref{fig:#1}}

\shorttitle{EM Cascade Emission from Neutrino-Coincident TDEs}
\shortauthors{Yuan and Winter}

\begin{document}
\begin{CJK*}{UTF8}{gbsn}

\title{Electromagnetic Cascade Emission from Neutrino-Coincident Tidal Disruption Events}
\correspondingauthor{Chengchao Yuan}
\email{chengchao.yuan@desy.de}

\correspondingauthor{Walter Winter}
\email{walter.winter@desy.de}

\author[0000-0003-0327-6136X]{Chengchao Yuan (袁成超)}\affil{Deutsches Elektronen-Synchrotron DESY, 
Platanenallee 6, 15738 Zeuthen, Germany}

\author[0000-0001-7062-0289X]{Walter Winter}\affil{Deutsches Elektronen-Synchrotron DESY, 
Platanenallee 6, 15738 Zeuthen, Germany}

\begin{abstract}
The potential association between Tidal Disruption Events (TDEs) and high-energy astrophysical neutrinos implies the acceleration of cosmic rays. These accelerated particles will initiate electromagnetic (EM) cascades spanning from keV to GeV energies by the processes related to neutrino production. We model the EM cascade and neutrino emissions by numerically solving the time-dependent transport equations and discuss the implications for AT2019dsg and AT2019fdr in the X-ray and $\gamma$-ray bands. We show that the $\gamma$-ray constraints from \emph{Fermi} can constrain the size of the radiation zone and the maximum energy of injected protons, and that the corresponding expected neutrino event numbers in follow-up searches are limited to be less than about 0.1. {Depending on the efficiency of $p\gamma$ interactions and the time at which the target photons peak}, the X-ray and $\gamma$-ray signals can be expected closer to the peak of the optical-ultraviolet (OUV) luminosity, or to the time of the neutrino production. 
 %whereas a time delay of the order of $p\gamma$ cooling time is expected in an extended radiation zone. 
% A multi-messenger diagnosis, incorporating the EM cascades, can provide valuable insights into the physical conditions of the particle acceleration and radiation zones, such as their magnetic fields and sizes.
\end{abstract}
%%%%%%%%%%%%%% Section 1: Introduction %%%%%%%%%%%%%
\section{Introduction}

Recent observations have revealed that TDEs can produce intense flares of radiation lasting from months to years, powered by infalling material from a tidally disrupted star.  
 Observationally, TDEs commonly exhibit thermal emissions predominantly in the OUV range, with sub-populations also observed in the X-ray and radio bands \citep[see, e.g.,][]{2020SSRv..216...85S,2020SSRv..216...81A,2021ApJ...908....4V}. The TDE catalog keeps rapidly expanding as more and more TDEs have been continuously identified by the Zwicky Transient Facility \citep[ZTF, e.g.,][]{2023ApJ...942....9H}.
 
 Three TDE candidates, AT2019dsg \citep{2021NatAs...5..510S}, AT2019fdr \citep{2022PhRvL.128v1101R} and AT2019aalc \citep{2021arXiv211109391V}, are potentially correlated to the three IceCube astrophysical neutrino events, IC191001A, IC200530A and IC19119A, respectively. Two of these neutrino associations (AT2019dsg and AT2019fdr) have been found by follow-up searches. It has been pointed out in \citet{2021arXiv211109391V} that these associations have a strong dust echo in common -- which is delayed re-processed radiation from the  OUV and X-ray ranges into the infrared (IR) range by surrounding dust; this has led to the identification of the third candidate (AT2019aalc).
 Remarkably, these three TDEs are located in the 90\% containment boxes of the corresponding neutrino events, confirmed in the recent catalog of IceCube neutrino tracks \citep{2023arXiv230401174A}, which consolidates the multi-messenger correlations. Apart from the strong dust echoes and OUV luminosities, AT2019dsg and AT2019fdr have been observed in X-rays, and the neutrinos have come delayed by half a year to a year with respect to the OUV peak.
 The neutrino counterparts of these three TDEs have a profound impact on TDE models and bring significant momentum to related researches since astrophysical neutrinos are primarily generated in hadronic processes requiring the acceleration of cosmic-ray primaries.

Extensive efforts have been made on theoretical modeling, simulations and observations to establish a unified physical picture for TDEs. In this picture, the accretion disks, sub-relativistic outflows or winds, and possibly a jet and a dust torus are typically included \citep[see, e.g.,][]{2018ApJ...859L..20D}. Motivated by the observation of the luminous jetted TDE Swift J1644+57 in 2011 \citep{2011Natur.476..421B}, relativistic jets have been proposed as one promising origin of the non-thermal emissions of TDEs \citep{2011PhRvD..84h1301W,2016PhRvD..93h3005W,2017MNRAS.469.1354D,2017PhRvD..95l3001L,2017ApJ...838....3S}. In the jetted models, the energy released is beamed in a narrow region where the power is sufficient high to explain the intense emissions. Another jetted TDE candidate, AT2022cmc \citep{2022Natur.612..430A}, has been recently reported to exhibit a very high non-thermal X-ray luminosity $\gtrsim10^{47}-10^{48}~\rm erg\ s^{-1}$, which is likely produced by a collimated jet with an opening angle $\theta_j\lesssim1^\circ$ \citep{2023NatAs...7...88P}. It was also pointed out that only around one percent of TDEs are expected to have relativistic jets; however, this small fraction could be consistent with the fraction of neutrino-emitting TDEs from diffuse flux considerations \citep{2023ApJ...948...42W}.

Apart from jets, quasi-isotropic emissions from accretion disks \citep{2019ApJ...886..114H}, wide-angle outflows/winds \citep{2020ApJ...904....4F} or tidal stream interactions \citep{2015ApJ...812L..39D,2019ApJ...886..114H} have been considered, particularly for TDEs that do not exhibit direct jet signatures, such as radio signals. The TDE jets, disks, and winds can be efficient cosmic-ray (CR) accelerators, potentially producing ultra-high-energy cosmic rays \citep[UHECRs,][]{2009ApJ...693..329F,2014arXiv1411.0704F,2017PhRvD..96f3007Z,2018A&A...616A.179G,2018NatSR...810828B}. Consequently, these sites may also serve as promising neutrino emitters. For AT2019dsg, various models such as jets, disks, corona and hidden winds and outflow-cloud interactions have been proposed to explain the neutrino counterparts \citep{2020PhRvD.102h3028L, 2021NatAs...5..472W, 2022MNRAS.514.4406W, 2020ApJ...902..108M, 2021NatAs...5..436H}. Similarly, for AT2019fdr, \cite{2022PhRvL.128v1101R} proposed the corona, hidden wind and jet models, whereas a disk model for all three neutrino-coincident TDEs has been proposed by \cite{2021arXiv211109391V}. Motivated by the absence of definitive evidence for jet signatures \citep[e.g.,][]{2022ApJ...927...74M}, \cite{2023ApJ...948...42W} have recently presented unified time-dependent interpretations of the neutrino emissions by considering three models (named M-IR, M-OUV, and M-X) where thermal IR, OUV and X-ray photons respectively dominate the neutrino productions depending on the maximally available proton acceleration energy. An alternative model, proposed by \cite{2022arXiv220911005Z}, has considered that the jets are choked inside the ejecta.

High-energy astrophysical neutrinos typically originate from charged pions produced in cosmic ray interactions with matter or radiation. While the charged pion decay chain also leads to electrons and positrons as decay products, neutral pions co-produced with the charged pions directly decay into gamma-rays. Therefore, a similar amount of energy is expected to be injected into neutrinos and the electromagnetic (EM) cascade in source, where
%The secondary electrons/positrons lose energy in synchrotron radiation and may up-scatter photons in inverse Compton processes, while the gamma-rays lead, due to the high densities, to electron-positron pair production with corresponding consecutive processes. 
it is a priori not clear where in energy the EM signatures dominate, and what the time-dependent behavior of the EM cascade is -- these questions require dedicated theoretical modeling, which we perform in this study. A joint analysis of the non-thermal X-ray and $\gamma$-ray observations with the neutrino detection will then provide further insights: The X-ray light curve for AT2019dsg was measured using Swift X-ray Telescope \citep[XRT,][]{2005SSRv..120..165B}, XMM-Newton \citep[XMM,][]{2001A&A...365L...1J}, and NICER \citep{2021NatAs...5..510S,2021MNRAS.504..792C}. As for AT2019fdr, upper limits were obtained from XRT and eROSITA \citep{2020SSRv..216...85S,2022PhRvL.128v1101R,2021A&A...647A...1P}. The \emph{Fermi} Large Area Telescope \citep{2009ApJ...697.1071A} did not detect significant $\gamma$-ray signals from both of them, resulting in a time-dependent up limit of $10^{-9}-10^{-8}~\rm GeV~cm^{-2}~s^{-1}$ \citep{2021NatAs...5..510S}.  Given the relatively incomplete X-ray and $\gamma$-ray datasets for AT2019aalc \citep{2021arXiv211109391V}, we choose AT2019dsg and AT2019fdr as two prototypes to study the implications of the EM cascades. 

We adopt the quasi-isotropic model proposed by \cite{2023ApJ...948...42W} and specifically study the time-dependent EM cascade emissions in radiation zones characterized generically by their sizes and the maximum energies of injected protons. Compared to \cite{2023ApJ...948...42W}, where the simulations were based on the NeuCosmA (Neutrinos from Cosmic Accelerators) code~\citep{2010ApJ...721..630H,2018A&A...611A.101B}, we change technology to self-consistently the electromagnetic cascade in-source. We use the AM$^3$ (Astrophysical Modeling with Multiple Messengers) software~\citep{2017ApJ...843..109G}, which has been successfully used for lepto-hadronic models of Active Galactic Nuclei blazars \citep[see, e.g.,][]{2019NatAs...3...88G,2019ApJ...874L..29R,2021ApJ...912...54R} and Gamma-Ray Bursts \citep{2022arXiv221200765R,2023ApJ...944L..34R}, and apply it for the first time to TDEs.
We address several key questions regarding EM cascade emissions in TDEs, e.g., what is the spectral shape of EM cascade emissions? What kind of observational features can be attributed to EM cascades? At what timescales and in which energy ranges can we expect to observe these EM cascades? What insights can be derived from X-ray and $\gamma$-ray observations, combined with neutrino counterparts, to constrain the parameter space of the model? For most of the results, we focus on the cases M-IR and M-OUV, where the thermal IR and OUV photons, respectively,  dominate the neutrino production. The M-X case, for which X-ray targets dominate, will only be included in some of the results as the qualitative conclusions are similar to M-IR.

The paper is organized as follows: in Sec. \ref{sec:TDE} we provide a brief description of the TDE models, particle interactions and numerical methods. EM cascade spectra and light curves for each TDE for the cases M-IR and M-OUV will be presented in Sec. \ref{sec:cascade}. In Sec. \ref{sec:constraints}, we show the $\gamma$-ray constraints on model parameters and predicted neutrino numbers for IceCube. We discuss our results and summarize our conclusions in Sec. \ref{sec:discussion} \& \ref{sec:summary}.

%%%%%%%%%%%%%% Section 2: TDE models %%%%%%%%%%%%%
\section{Model description}\label{sec:TDE}

In this section, we describe the TDE model in this work, and the numerical approaches employed for computing neutrino and EM cascade emissions. 

\subsection{Overview}

We focus on the description of a spherical radiation zone of radius $R$ leading to quasi-isotropic emission of neutrinos and photons, following  \citet{2023ApJ...948...42W}. The acceleration is assumed to occur inside this region at a radius $R_{\rm acc} \lesssim R$.\footnote{This relationship applies to models M-OUV and M-IR (see below), whereas $R_{\rm acc} \sim R$ for model M-X.} We do not simulate the accelerator explicitly, we instead parameterize the acceleration zone by the maximal proton energy $E_{p,\mathrm{max}}$ and the  dissipation efficiency $\varepsilon_{\rm diss}$, which describes the conversion efficiency from the mass accretion rate into non-thermal protons. Because of the quasi-isotropic emission, large values of $\varepsilon_{\rm diss} \simeq 0.05-0.2$ are required to obtain reasonable neutrino event rates; we will use a value on the upper end of this range to saturate the EM cascade bounds. Note that its precise value is not so relevant for the conclusions of this work, as it scales the neutrino and EM emissions in the same way. The accelerated protons are assumed to be magnetically confined in magnetic fields of strength $B$ as long as the Larmor radius is smaller than the size of the region, but they can diffuse out in the Bohm limit. 

Protons in the radiation zone may interact with IR, OUV, and X-ray radiation parameterized as thermal spectra with temperatures $T_{\mathrm{IR}}$, $T_{\mathrm{OUV}}$, and $T_{\mathrm{X}}$, respectively; the higher the target photon energy is, the lower is the required $E_{p,\mathrm{max}}$. Consequently, the models are labeled M-X (lowest $E_{p,\mathrm{max}}$, interactions with X-ray target only), M-OUV (intermediate $E_{p,\mathrm{max}}$, interactions dominated by OUV target), and M-IR (highest $E_{p,\mathrm{max}}$, interactions with IR target possible). Our interaction targets are based solely on observations, which means that the do not add any other theoretically motivated ingredients which have not been directly observed. {For the scenarios M-OUV and M-IR, the radii of radiation zones are determined by the regions of respective isotropized thermal target photons, whereas the acceleration is expected to take place somewhere inside these regions. For instance, we choose $R\sim10^{15}~\rm cm$ (which is close to OUV photons) and $R\sim10^{17}~\rm cm$ (which is close to IR photons originated from the dust torus) for the M-OUV and M-IR scenarios, respectively.} Our models are fully time-dependent, taking into account the time-dependence of the proton injections assumed to follow the OUV light curves, and the time-dependence of the target photons. This is illustrated in \figu{luminosities}, where the time-dependent proton injection lumimosities and the time evolutions of the photon targets are shown for AT2019dsg and AT2019fdr; further model details are listed in Tab.~\ref{tab:params}. It is especially noteworthy that the IR photons are assumed to be be produced by the dust as so-called ``dust echoes'', where dust heated by X-ray and OUV radiation may re-emit IR photons \citep{2016ApJ...829...19V,2021arXiv211109391V,2022PhRvL.128v1101R}. We further on use the dust echo IR light curves obtained by \cite{2023ApJ...948...42W}.

We focus on two scenarios, M-IR and M-OUV first, as they are representative for our results. However, we will later consider extended radiation zones (e.g., located close to the dust radius) and compact radiation zones (e.g., located close to the OUV-BB), respectively, as we will use $R$ as a free parameter in Sec. \ref{sec:constraints}. Further model details are given in the following subsections.

\subsection{Proton injection}
\label{subsec:TDE_proton}

\begin{table*}
    \centering
    \caption{Observational and TDE modeling parameters for AT2019dsg and AT2019fdr. In all scenarios, the universal values of energy dissipation efficiency $\varepsilon_{\rm diss}=0.2$ and magnetic field strength $B=0.1~\rm G$ are used.} 
    \begin{ruledtabular}
    \begin{tabular}{c|cc|cc}
        &\multicolumn{2}{c|}{\bf AT2019dsg$^a$

        }&\multicolumn{2}{c}{\bf AT2019fdr$^b$
        
        }\\
        
        &\multicolumn{2}{c|}{$z=0.051,~M=5\times10^6M_\odot$, $t_{\rm dyn}=670~\rm d$}&\multicolumn{2}{c}{$z=0.267,~M=1.3\times10^7M_\odot,~t_{\rm dyn}=1730~\rm d$}\\
        %\specialrule{.1em}{.05em}{.05em} 
        \hline
        
        $k_BT_{\rm X,~OUV,~IR}$&\multicolumn{2}{c|}{72 eV, 3.4 eV, 0.16 eV}&\multicolumn{2}{c}{56 eV, 1.2 eV, 0.14 eV}\\
        
        $E_{\nu}
        $&\multicolumn{2}{c|}{217 TeV (IC191001A)}&\multicolumn{2}{c}{82 TeV (IC200530A)}\\
        
        $t_\nu-t_{\rm pk}$&\multicolumn{2}{c|}{154~\rm d}&\multicolumn{2}{c}{324~\rm d}\\
        
        $N_{\nu}(\rm GFU)^c$
        &\multicolumn{2}{c|}{$0.008-0.76$}&\multicolumn{2}{c}{$0.007-0.13$}\\
        
        \hline
        Scenario & M-IR & M-OUV & M-IR & M-OUV\\
         \hline
         
        $R$ [cm]  &$5.0\times10^{16}$ &$5.0\times10^{14}$ &$2.5\times10^{17}$&5.0$\times10^{15}$\\
        
        $E_{p,\rm max}$ [GeV]  &$5.0\times10^9$ &$1.0\times10^{8}$ &$5.0\times10^9$&1.0$\times10^{8}$\\
     \end{tabular}
    \end{ruledtabular}

    \begin{tablenotes}
			\item[(1)] $^a$AT2019dsg data references: redshift $z$, expected neutrino number via IceCube GFU searches $N_{\nu}$(GFU), $T_{\rm OUV}$ and $T_{\rm X}$ \citep{2021NatAs...5..510S}; SMBH mass $M$ \citep{2021arXiv211109391V}; peak time of OUV light curve $t_{\rm pk}$ \citep{2021NatAs...5..510S}; Neutrino energy $E_\nu$ \citep{2019GCN.25913....1I}; $T_{\rm IR}$ \citep{2023ApJ...948...42W}.
   
			\item[(2)] $^b$AT2019fdr data references: $z$, $t_{\rm pk}$, $N_{\nu}$(GFU), $T_{\rm OUV}$, $T_{\rm X}$ and $T_{\rm IR}$  \citep{2022PhRvL.128v1101R}; $M$ \citep{2021arXiv211109391V};  $E_\nu$ \citep{2019GCN.26258....1I}.

            \item[(3)] $^c$Expected neutrino number from IceCube gamma-ray follow up (GFU) searches. 
		\end{tablenotes}
    \label{tab:params}
\end{table*}

\begin{figure*}\centering
\includegraphics[width=0.49\textwidth]{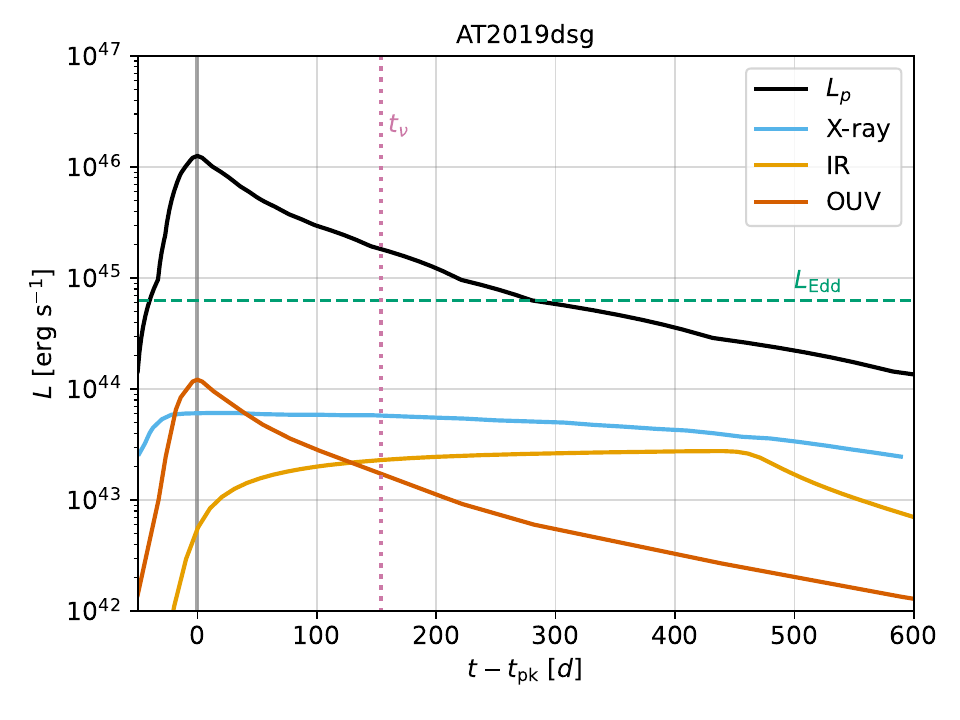}
\includegraphics[width=0.49\textwidth]{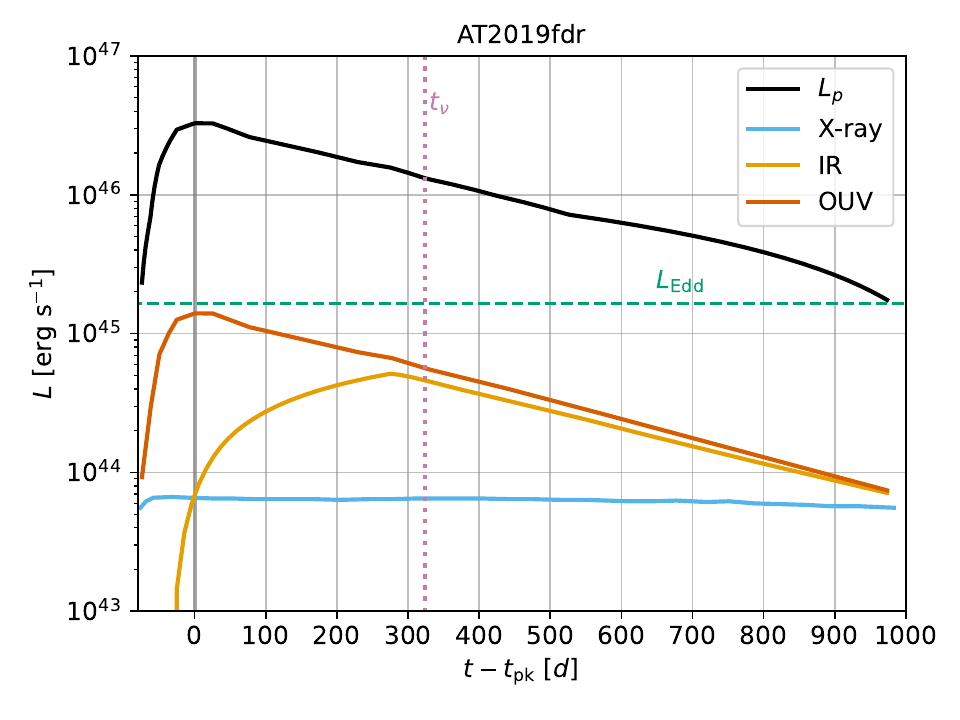}
\caption{Time-dependent evolution of IR, OUV, X-ray, and injected proton luminosities, shown as the orange, red, blue, and black solid curves respectively. The dissipation efficiency $\varepsilon_{\rm diss}=L_p/(\dot Mc^2)\simeq0.2$ is used to obtain the proton injection luminosity $L_p$ for AT2019dsg (left panel) and AT2019fdr (right panel). The horizontal green dashed curves represents the Eddington luminosity $L_{\rm Edd}$. The OUV luminosity peaks ($t_{\rm pk}$) and neutrino detection times ($t_\nu-t_{\rm pk}$) are illustrated as the vertical gray solid and magenta dotted curves, respectively.}
\label{fig:luminosities}
\end{figure*}

 Possible locations for the acceleration zone include the accretion disk corona, relativistic jets, or the disk winds; correspondingly, $R_{\rm acc}$ may vary significantly. Near the X-ray photosphere or the disk corona, the size can be $\sim(10-30)R_{s}$ \citep[e.g.,][]{2020ApJ...902..108M,2021arXiv211109391V}, where $R_s=2GM/c^2$ is the Schwarzschild radius of a SMBH with mass $M$. Inside the jet, particle acceleration is likely to occur at a radius of a few hundred to 1000 $R_s$ \citep[e.g.,][]{2018ApJ...859L..20D}, depending on the location of the shocks. Alternatively, if particle acceleration in the wind is considered, $R_{\rm acc}$ can extend to the boundary of the wind envelope, e.g., $R_{\rm acc}\sim10^{16}-10^{17}~\rm cm$ \citep{2021NatAs...5..510S}, in proximity to the dust radius. 

In this work, we focus on the description of the radiation zone characterized by the radius $R\gtrsim R_{\rm acc}$ without explicitly specifying the particle acceleration site, indicating that the physical conditions of the radiation and acceleration regions, such as magnetic field strength, may not necessarily be identical. Note that even though the proton emission from the accelerator may be anisotropic, the magnetic field in the radiation zone may isotropize the protons in the calorimetric limit; an example would be an off-axis jet as accelerator in a large radiation zone (model M-IR).   Using an acceleration rate $t_{\rm acc}^{-1}=\eta_{\rm acc} (c/R_{L,\rm acc})\lesssim c/R_{L,\rm acc}$ \citep{1984ARA&A..22..425H}, the protons could be accelerated to $10^{8}-10^9~\rm GeV$ near the accretion disk corona ($R_{\rm acc}\lesssim10^{14}~\rm cm$) with a strong magnetic field strength or in the extended disk wind ($R_{\rm acc}\sim10^{17}~\rm cm$), where $R_{L,\rm acc}=E_p/(eB_{\rm acc})$ is the Larmor radius of protons in the acceleration zone characterized by the magnetic field strength $B_{\rm acc}$, $\eta_{\rm acc}\lesssim1$ is the acceleration efficiency and $t_{\rm acc}$ is the acceleration time required to energize protons to $E_p$ \citep[see discussion in][]{2023ApJ...948...42W}. 

We adopt an energy-differential power-law proton injection spectrum (per unit volume per unit time)  $Q_p(E_p,t)\propto E_p^{-2}\exp(-E_p/E_{p,\rm max})$ where $E_{p,\rm max}$ is the maximum energy of injected proton spectrum and is treated as a free parameter. The spectrum can be normalized via $\int E_pQ_pdE_p=L_p/V$, where $L_p$ denotes the time-dependent proton luminosity and $V=4\pi R^3/3$ is the volume of the radiation region. Typically, the minimum proton energy can be written as $E_{p,\rm min}=\Gamma_{\rm rel}m_pc^2$, where $\Gamma_{\rm rel}$ is the relative Lorentz factor between the acceleration site and the radiation zone. Since the absence of the specification of the  acceleration site, the conservative value of $E_{p,\rm min}=1~\rm GeV$ is used. As depicted by the red curves in Fig. \ref{fig:luminosities}, OUV observations reveal that TDEs generically exhibit characteristic light curves, with an initial rapid increase and a subsequent slower decay ($\propto t^{-5/3}$) that is consistent with the predicted mass fallback rate. This behavior suggests the OUV luminosity ($L_{\rm OUV}$) reflects the accretion history. Henceforth, following the methodology in \cite{2023ApJ...948...42W}, we assume that a fraction of the accreted energy onto the SMBH is dissipated into protons and the accretion rate aligns with the observed temporal evolution of OUV light curves. We write down $L_{p}$ in terms of the Eddington accretion rate $\dot M_{\rm Edd}$, which represents the accretion rate required for a black hole to radiate at the Eddington luminosity $L_{\rm Edd}\simeq1.3\times10^{45}{~\rm erg~s^{-1}}M/(10^7M_\odot)$, as
\begin{equation}
    L_p=\varepsilon_{\rm diss}\dot Mc^2=\varepsilon_{\rm diss}\zeta \dot M_{\rm Edd}c^2\frac{L_{\rm OUV}}{L_{\rm OUV}(t_{\rm pk})}.
    \label{equ:CR_luminosity}
\end{equation}
In this expression, $t_{\rm pk}$ is the time when OUV light curve reaches its peak $L_{\rm OUV}(t_{\rm pk})$, and $\zeta\equiv \dot M(t_{\rm pk})/\dot M_{\rm Edd}$ is the ratio of the peak accretion rate to the Eddington accretion rate $\dot M_{\rm Edd}$. Generally, $\dot M_{\rm Edd}$ can be estimated from  the radiation efficiency  $\eta_{\rm rad}\sim0.01-0.1$ \citep[e.g.,][]{2015MNRAS.454L...6M}, that represents the fraction of accreted energy reprocessed into radiation energy at the level of $L_{\rm Edd }$,  via $\dot M_{\rm Edd}=L_{\rm Edd}/(\eta_{\rm rad}c^2)$.  Considering the super-Eddington accretion nature of TDEs during the peak accretion phase, e.g., $\zeta\sim10-100$ \citep[e.g.,][]{2018ApJ...859L..20D}, we establish the relationship, $\dot M(t_{\rm pk})=\zeta\dot M_{\rm Edd}=(\zeta/\eta_{\rm rad}) L_{\rm Edd}/c^2$. We adopt $\zeta/\eta_{\rm rad}=100$ in the following analysis; this value can be also justified by comparing the time-integrated accretion mass with the the mass of disrupted stars \citep{2023ApJ...948...42W}. Under these parameterizations, the thick black curves in the left and right panels of Fig. \ref{fig:luminosities} depict the injected proton luminosity for AT2019dsg and AT2019fdr. We show also the Eddington luminosity as green dashed lines for reference.

\subsection{Interactions and numerical methods}
\label{subsec:target_photon_interaction}

In the radiation zone, neutrinos can be generated through photomeson ($p\gamma$) and hadronuclear ($pp$) interactions. In the $p\gamma$ process, the quasi-thermal X-ray, OUV, and IR emissions serve as the target photon fields. With the time-dependent luminosity $L$ obtained from direct observations of the IR/OUV/X-ray components, we calculate the target photon density through 
\begin{equation}
    \int \varepsilon_\gamma n(\varepsilon_\gamma)d\varepsilon_\gamma=\frac{L}{4\pi R^2c},
    \label{equ:target_photon_norm}
\end{equation}
where 
\begin{equation}
    n\propto\frac{\varepsilon_\gamma^2}{\exp\left(\frac{\varepsilon_\gamma}{k_BT}\right)-1}
    \label{equ:target_photon_spec}
\end{equation}
is the black body spectrum (in the units of $\rm eV^{-1}~cm^{-3}$) and $T$ is the characteristic temperature determined by fitting the observed spectra. For OUV components, the photon density can be straightforwardly obtained from Eqs. (\ref{equ:target_photon_norm}) and (\ref{equ:target_photon_spec}) using the observed $L_{\rm OUV }$ and $T_{\rm OUV}$. In the context of X-ray photons originated from accretion disks, we assume that the in-source X-ray luminosity remains relatively stable \citep[e.g.,][]{2020ApJ...897...80W}. The time-dependent fluctuations observed in X-ray observations could be attributed to absorption and geometrical obscuration. Therefore, we consider the high-level observed luminosity as the representative X-ray luminosity within the radiation zone, as described by \cite{2023ApJ...948...42W}. Note that this assumption is only relevant in Sec.~\ref{sec:constraints} for small enough proton energies.
Regarding the IR emissions measured subsequent to the OUV peak, the dust echo interpretation suggests that echo IR photons are generated by dust located at a distance of approximately $R_{\rm IR}\sim10^{16}-10^{17}~\rm cm$. The dust heated by the X-ray and OUV radiation will re-emit IR photons with a temperature of approximately $k_BT_{\rm IR}\simeq 0.16~\rm eV$ \citep{2016ApJ...829...19V,2021arXiv211109391V,2022PhRvL.128v1101R}. Additionally, the deflection of these photons with respect to the line of sight explains the observed time delay. \cite{2022PhRvL.128v1101R} have demonstrated that the IR light curve can be explained by convolving the OUV luminosity with a normalized box function. Using this method, \cite{2023ApJ...948...42W} performed a least-square fitting of the IR observations and obtained the IR light curves for AT2019dsg and AT2019fdr. Building upon the work of \cite{2023ApJ...948...42W}, we incorporate the IR light curves presented in their study directly into our modeling of the target photon density. Figure \ref{fig:luminosities} illustrates the time dependence of IR ($L_{\rm IR}$; orange lines), OUV ($L_{\rm OUV}$; red lines), and X-ray ($L_X$; blue lines) luminosities. The vertical purple dotted lines represent the time of neutrino detection, $t_{\nu}-t_{\rm pk}$. The time axes in this figure are defined with the OUV luminosity peak time $t_{\rm pk}$ as the origin. The IR, OUV, and X-ray target photon densities are determined by \equ{target_photon_norm} and \equ{target_photon_spec}, using the respective luminosities. The measured temperatures of the IR, OUV, and X-ray black body emissions are provided in Tab. \ref{tab:params}. 

While propagating inside the radiation zone, the injected protons will also interact with the thermal protons in the sub-relativistic outflow or wind. Given the wind mass rate $\dot M_{w}=\eta_{w}\dot M$ \citep{2018ApJ...859L..20D,2021ApJ...911L..15Y},  where $\eta_w \equiv \dot M_w/\dot M$ is the fraction of accreted mass converted to wind, we estimate the average target proton number density $n_w\sim\eta_w\dot M/(4\pi m_pv_wR^2)$ and the $pp$ interaction rate  
\begin{equation}
    t_{pp}^{-1}\simeq cn_w\sigma_{pp}\sim \frac{1}{4\pi}\sigma_{pp}\beta_w^{-1}\eta_w\frac{\dot M}{R^2m_p} \, .
\end{equation}
Here $v_w=\beta_w c\sim\mathcal O(0.1c)$ is the wind velocity and $\sigma_{pp}$ is the cross section for inelastic $pp$ collisions. For super-Eddington accretors, $\eta_w$ typical ranges from $10^{-2}$ to $10^{-1}$ in the case of the efficient magnetically arrested disks \citep[see e.g.,][]{2009PASJ...61L...7O,2019ApJ...880...67J,2020PhRvD.102h3013Y}. We will demonstrate that the $p\gamma$ process is significantly more efficient than $pp$, even if an optimistic value of $\eta_w\sim0.1-0.2$ is used.

We also consider the electromagnetic cascade emission resulting from the secondary particles generated by the $p\gamma$ and $pp$ processes. Charged pions produced from $p\gamma$ and $pp$ interactions decay, yielding neutrinos, electrons, and positrons, while neutral pions directly decay into photons (denoted as `$\pi^0$ decay'). The energy deposited in the secondary electrons and positrons can then be redistributed to the radiation field through synchrotron (SY) and inverse Compton (IC) radiation in the presence of magnetic fields. We denote this process as `$pp/p\gamma$-SY/IC'.  As a competing channel to the $p\gamma$ process, protons can also lose energy through Bethe-Heitler (BH) interactions, leading to the production of electron-positron ($e^\pm$) pairs. Additionally, the $\gamma\gamma$ attenuatios between the EM cascade emissions and the abundant thermal IR, OUV and X-ray photons give rise to $e^\pm$ pairs. These $e^\pm$ pairs from BH and $\gamma\gamma$ processes are expected to contribute to the EM cascade radiation field, represented as `BH-SY/IC' and `$\gamma\gamma$-SY/IC' respectively. 

\begin{figure*}
    \centering
    \includegraphics[width=0.49\textwidth]{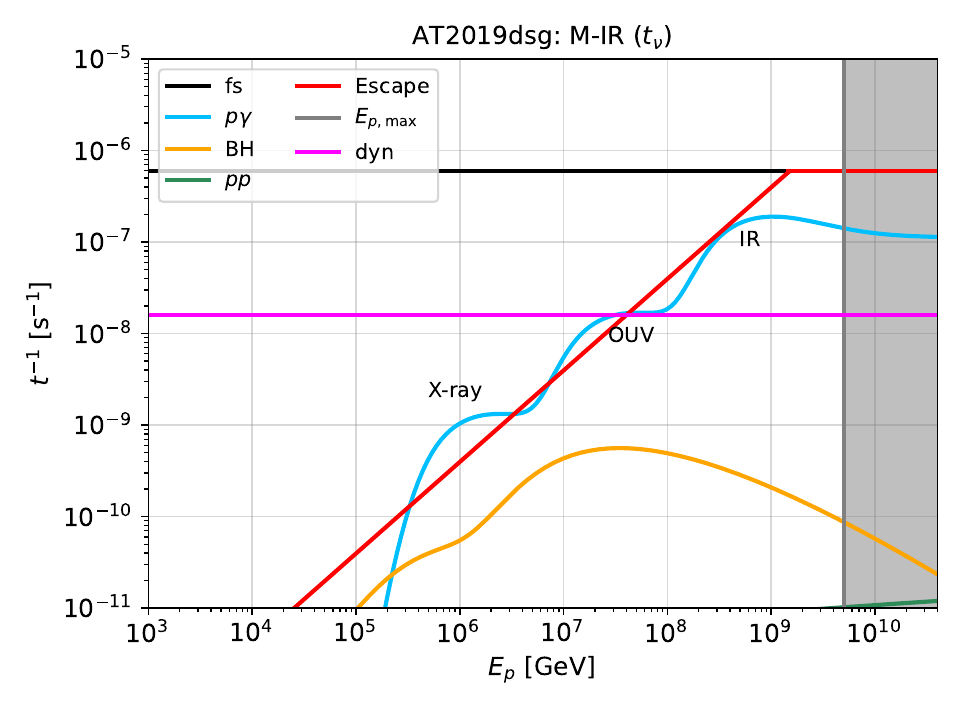}
    \includegraphics[width=0.49\textwidth]{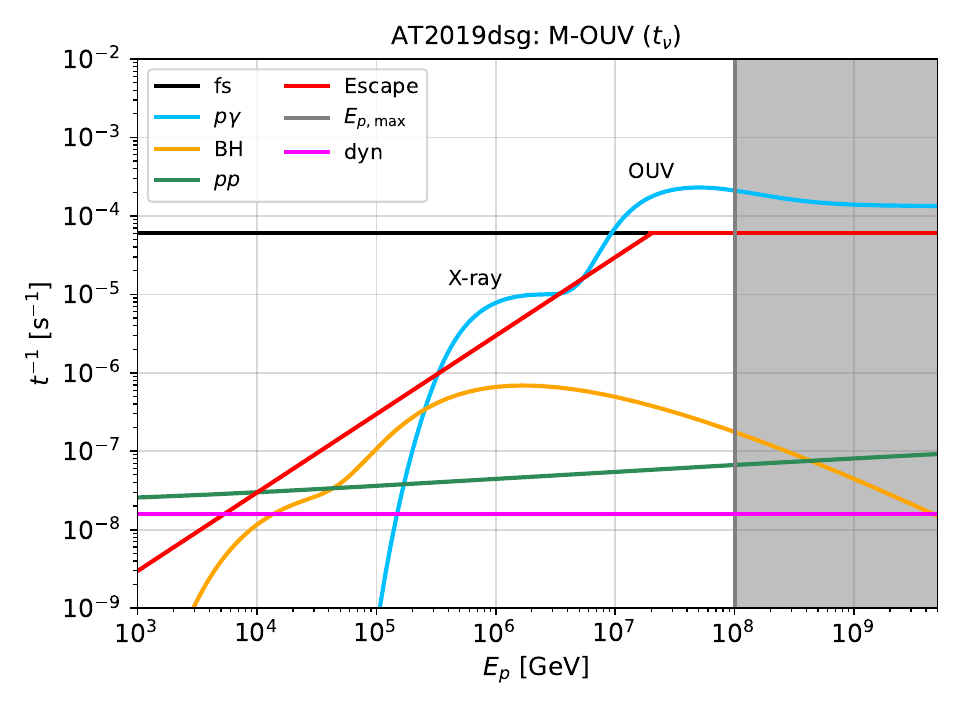}

    \caption{Proton interaction rates at neutrino detection time $t_\nu$ for AT2019dsg, as a function of proton energy (in the observer's frame). The left and right panels correspond to the M-IR and M-OUV scenarios. In each figure, the $p\gamma$, $pp$ with $\eta_w=0.2$, Bethe-Heitler (BH), escape, and dynamical rates are shown as the blue, green, orange, red, and magenta curves. The horizontal curves depict the free streaming rate $t_{\rm fs}^{-1}=(R/c)^{-1}$, whereas the vertical gray curves show the maximum injected proton energies $E_{p,\rm max}$. }
    \label{fig:timescales}
\end{figure*}

To obtain the neutrino and EM cascade spectra, we numerically solve the coupled time-dependent transport equations for all relevant particle species
\begin{equation}
\frac{\partial n_i}{\partial t}=Q_{i,\rm ext}+\sum_k Q_{k\rightarrow i}-\frac{\partial}{\partial E}(\dot E\cdot n_i)-\frac{n_i}{t_{i,\rm esc}}
\label{equ:transport_eq}
\end{equation}
in a self-consistent way using AM$^3$ software \citep[e.g.,][]{2017ApJ...843..109G}. 
In \equ{transport_eq}, $i$ is the label of the particle, $n_i\equiv dN_i/dVdE$ is the in-source particle density differential in energy and volume, $N_i$ is the total number, $Q_{i,\rm ext}$ is the external particle injection rate, e.g., $Q_{p,\rm ext}=Q_p(E_p,t)$, $\sum_{k}Q_{k\rightarrow i}$ represents the in-source particle injection from other radiation processes, $\dot E=dE/dt$ is the total energy loss rate, and $t_{i,\rm esc}$ is the particle escape time. Since we focus on the EM cascade emission induced by accelerated protons and do not consider the primary leptonic loading in the acceleration zone, the external electron/positron injection rate is set to zero, e.g., $Q_{e,\rm ext}=0$. The influence of primary electron injections will be discussed in Sec.~\ref{sec:discussion}. 

Regarding the particle escape times, the photons, neutrinos, and neutrons produced in $p\gamma$ processes escape the radiation zone with the free-streaming time $t_{\rm esc}=t_{\rm fs}=R/c$, whereas the escape rate of charged particles is determined by the diffusion rate. We use as proton escape rate $t_{p,\rm esc}^{-1}=\min[t_{\rm fs}^{-1},t_{p,\rm diff}^{-1}]$~\citep{2013ApJ...768..186B}, where $t_{p,\rm diff}^{-1}\equiv D/R^2$ is the diffusion rate, $D\sim R_Lc$ is the diffusion coefficient in Bohm limit, and $R_L=E_p/(eB)$ is the Larmor radius of protons with energy $E_p$ in the magnetic field $B$. For simplicity, we choose the magnetic field strength $B=0.1$ G as the fiducial value for both AT2019dsg and AT2019fdr as in \cite{2023ApJ...948...42W}. This choice of magnetic field is consistent with the case of outflows from AGNs and is sufficiently strong to confine protons with energies $E_p\gtrsim 1$ PeV in the radiation zone. We will investigate the impacts on neutrino and EM cascade emissions by varying $B$ from 0.1 G to 1.0 G in Sec.~\ref{sec:constraints}.

We have checked the consistency of our results with the earlier computations in \citet{2023ApJ...948...42W} using NeuCosmA. While our results overall agree very well, we find small differences e.g. in the shapes of the neutrino spectra at the level expected from different softwares using different solver and particle interaction implementations  \citep[see e.g.][]{2022icrc.confE.979C}; for example, these come from the different implementations of photohadronic interactions: while AM$^3$ uses \citet{2010ApJ...721..630H}, NeuCosmA is based on \citet{2018A&A...611A.101B}.

\subsection{M-IR and M-OUV scenarios}\label{subsec:IR_OUV_scenarios}

Based on the formulation described before, the physical properties of the radiation zone can be characterized by two parameters: $R$ and $E_{p,\rm max}$, assuming a fixed magnetic field strength of $B=0.1$ G. We investigate the EM cascade emissions of two scenarios, M-IR and M-OUV, for AT2019dsg and AT2019fdr. The corresponding parameter sets are listed in Tab. \ref{tab:params}.

\emph{M-IR scenario.} As for $p\gamma$ process, the proton threshold energy is
\begin{equation}
    E_{p,\rm th}=\frac{m_\pi(2m_p+m_\pi)c^4}{4E_\gamma}\simeq 7.1\times10^{8}~{\rm GeV}\left(\frac{E_{\gamma}}{0.1~\rm eV}\right)^{-1},
    \label{equ:pgamma}
\end{equation}
where $E_\gamma$ is target photon energy and $m_\pi\simeq140~\rm MeV/c^2$ is the pion mass.
We choose $R=5\times10^{16}~\rm cm\lesssim R_{\rm IR}$ and $E_{p,\rm max}=5.0\times10^9~{\rm GeV}\gtrsim E_{p,\rm th}$ for the M-IR scenario of AT2019dsg to ensure that $p\gamma$ interactions with thermal IR photons with energy $E_{\gamma,\rm IR}=k_BT_{\rm IR}\simeq0.16~\rm eV$ dominate the neutrino production. The left panel of Fig. \ref{fig:timescales} shows the proton interaction rates at neutrino detection time $t_\nu$ due to $p\gamma$ ($t_{p\gamma}^{-1}$, blue curve), $pp$ ($t_{pp}^{-1}$ with $\eta_w=0.2$, green line) and BH ($t_{\rm BH}^{-1}$, orange curve) interactions. The horizontal black line is the particle free-streaming rate $t_{\rm fs}^{-1}$, whereas the magenta horizontal line depicts the reciprocal of the dynamic time, e.g., $t_{\rm dyn}^{-1}$. We define the dynamic time ($t_{\rm dyn}$) as the duration of the super-Eddington phase, e.g., $\dot Mc^2>L_{\rm Edd}$. The energy-dependent escape rate is illustrated as the red line. The vertical gray line and the shaded region describe the maximum proton energy $E_{p,\rm max}$.

From this figure, we can deduce that the system is $p\gamma$ optically thin, indicated by $\tau_{p\gamma}^{\rm fs}=t_{p\gamma}^{-1}/t_{\rm fs}^{-1}<1$, but nearly calorimetric at ultra-high energies, as $\tau_{p\gamma}^{\rm cal}=t_{p\gamma}^{-1}/t_{\rm esc}^{-1}\sim1$ around $10^{9}$ GeV. This means that in the M-IR scenario, the TDE can be a promising very-high-energy neutrino emitter, but a time delay in the scale of $t_{p\gamma}$ of the $p\gamma/pp-$SY/IC cascade channel is typically expected. Furthermore, the $p\gamma$ process dominates the neutrino production as $t_{p\gamma}^{-1}\gg t_{pp}^{-1}$.

\emph{M-OUV scenario.} According to \equ{pgamma}, the proton threshold energy decreases with increasing energy of the target photon. We therefore consider a compact radiation zone with $R=5\times10^{14}~\rm cm\ll R_{\rm IR}$ and a relatively lower maximum proton energy $E_{p,\rm max}=1.0\times10^{8}~\rm GeV$ for the M-OUV scenario of AT2019dsg. In this case, $p\gamma$ interactions with OUV target photons dominate the neutrino flux. As shown in the right panel of Fig. \ref{fig:timescales}, the IR bump in the $p\gamma$ interaction rate (blue curve) is suppressed due to the higher density of OUV photons ($n_{\rm OUV}\propto R^{-2}$) compared to IR photons ($n_{\rm IR}\propto R_{\rm IR}^{-2}$). Notably, in this scenario, the source can be $p\gamma$ optically thick from $t_{\rm pk}$ to $t_{\nu}$, as indicated by the right panel of Fig. \ref{fig:timescales}, where $\tau_{p\gamma}^{\rm fs}=t_{p\gamma}^{-1}/t_{\rm fs}^{-1}>1$ is satisfied at $t_\nu$. The $p\gamma$ interactions can be much faster at $t_{\rm pk}$ where the OUV luminosity is approximately one order of magnitude higher (see the red curve in the left panel of Fig. \ref{fig:luminosities}). 

We use AT2019dsg as an example to illustrate the specifications of the M-IR and M-OUV scenarios. Similar conclusions can be drawn for AT2019fdr using the parameters listed in Tab. \ref{tab:params}.

\begin{figure*}[tp]\centering
\includegraphics[width=0.49\textwidth]{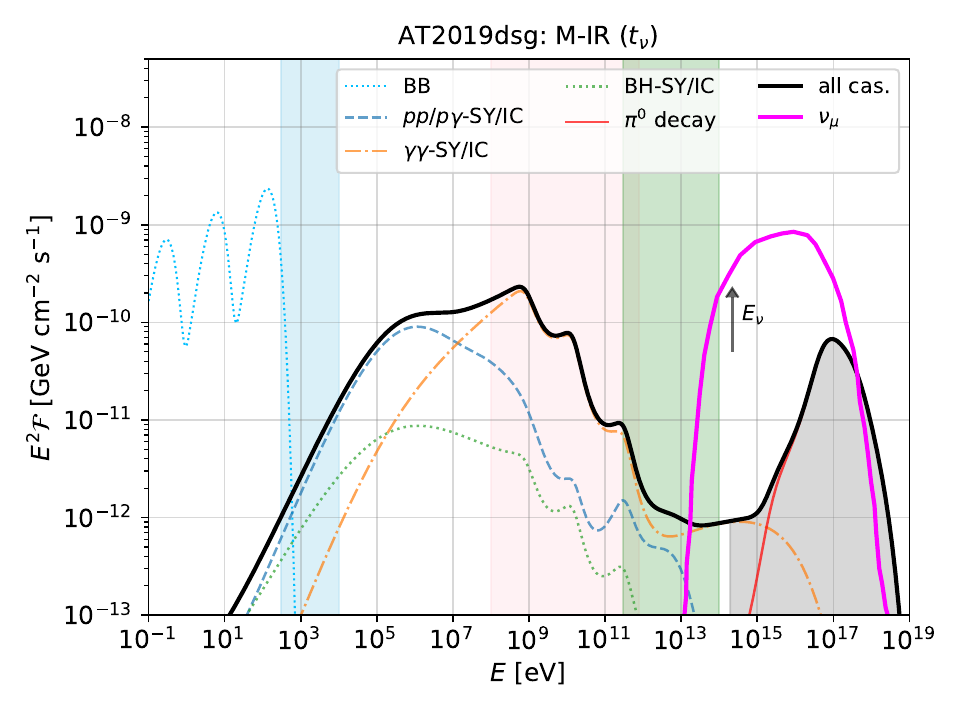}
\includegraphics[width=0.49\textwidth]{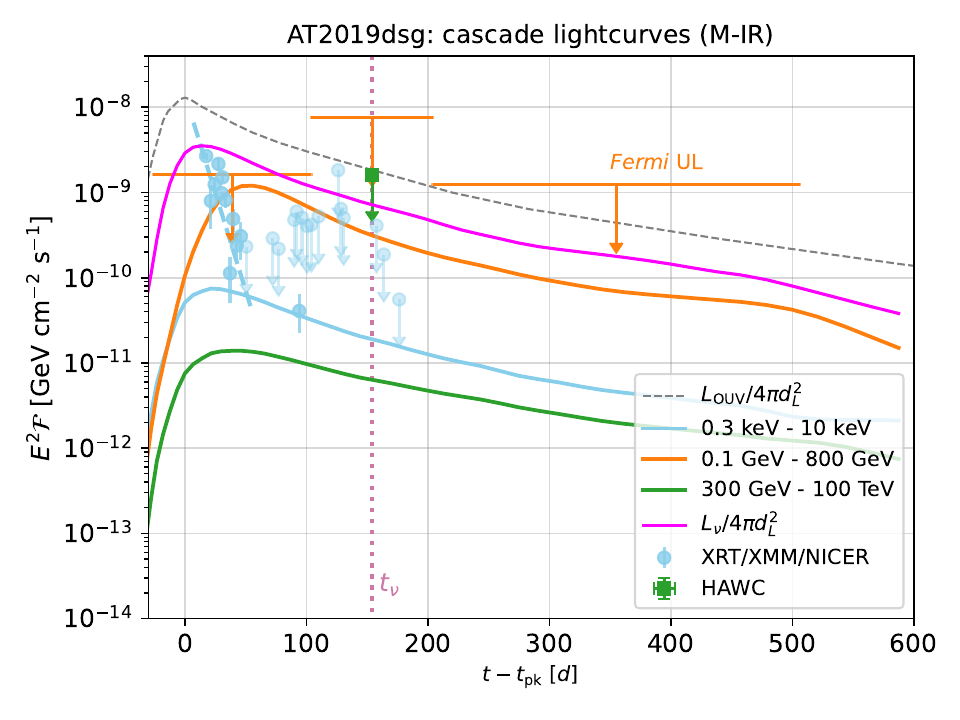}
\includegraphics[width=0.49\textwidth]{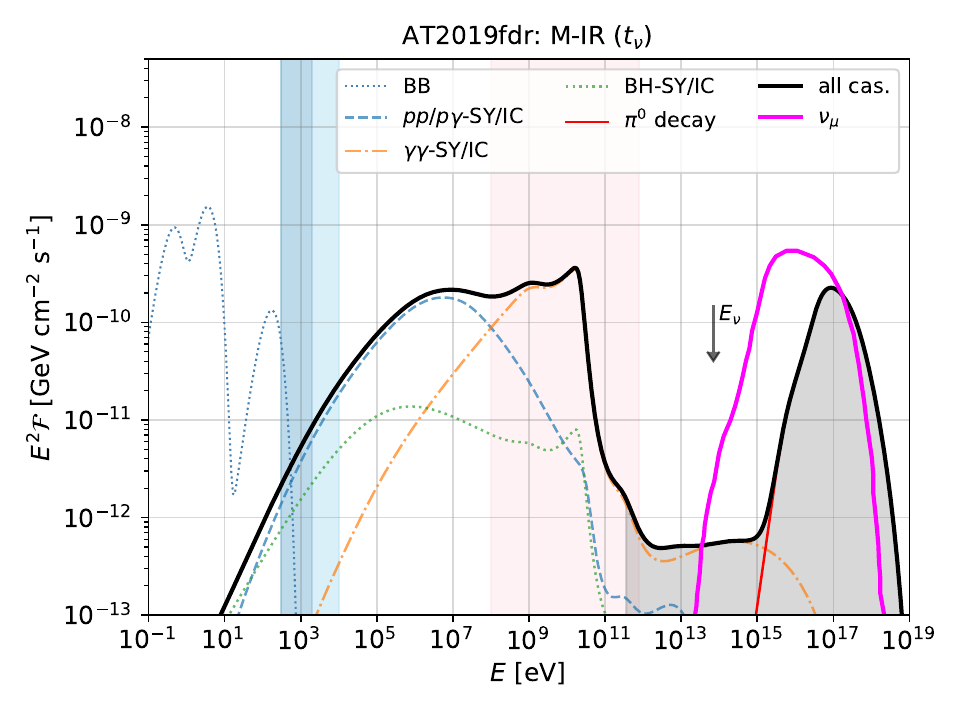}
\includegraphics[width=0.49\textwidth]{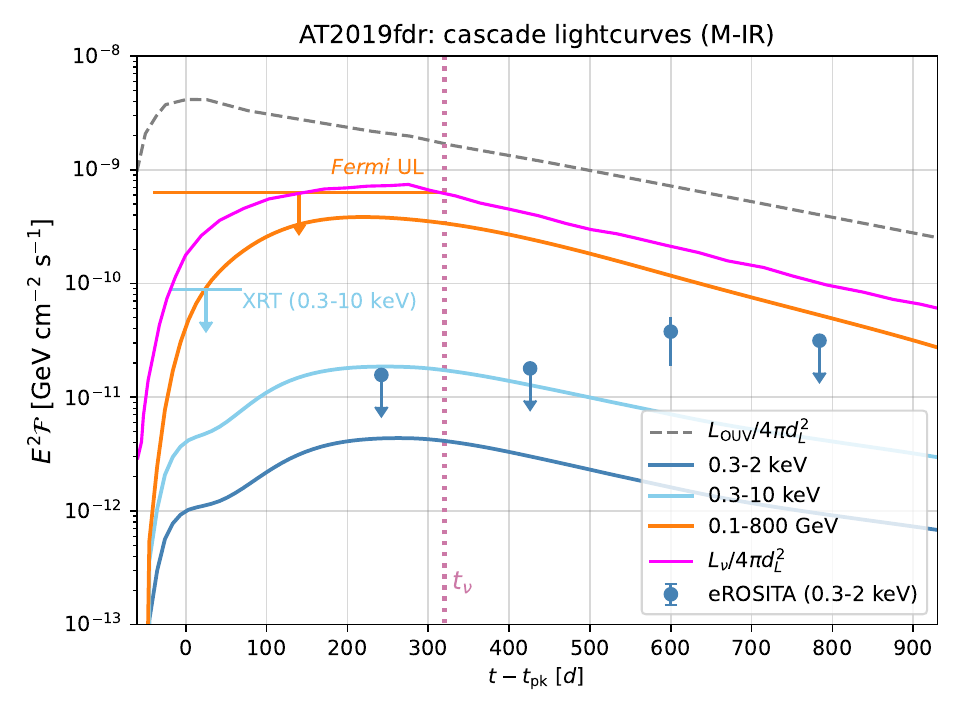}
\caption{{\bf Left column:} Electromagnetic cascade  and neutrino spectra at the time of the neutrino observation $t_\nu$. The black and magenta curves show the spectra of EM cascade and single-flavor neutrino emissions. The black body IR, OUV and X-ray spectra are depicted as the thin cyan dotted curves. The thin blue dashed, orange dash-dotted, green dotted, and red curves represent the components contributing to the EM cascade. The blue, red, and green shaded areas depict the energy ranges of XRT, \emph{Fermi}-LAT, and HAWC observations, whereas the gray-shaded areas indicate where $\gamma\gamma$ attenuation with CMB and EBL prohibits the observation of the emitted gamma rays.
{\bf Right column:}  Neutrino, $\gamma$-ray and X-ray light curves.   The the magenta, green, orange, and blue solid curves illustrate the single-flavor neutrino ($L_\nu/4\pi d_L^2$), very-high-energy (VHE) $\gamma$-ray (0.3 - 100 TeV), $\gamma$-ray (0.1 - 800 GeV), and X-ray light curves, respectively. The flux up limits from HAWC, \emph{Fermi}, and XRT/XMM/NICER or eROSITA \citep{2021NatAs...5..510S,2021MNRAS.504..792C,2021A&A...647A...1P,2022PhRvL.128v1101R} are shown as the green, orange, and blue points, respectively. Furthermore, the neutrino detection times are marked as the vertical magenta dotted curves. The OUV fluxes are also plotted as the gray dashed curves for reference. Light blue shaded areas and curves correspond to XRT (0.3 - 10 keV), whereas dark blue shaded areas and curves to eROSITA (0.3 - 2 keV).
The upper and lower rows correspond to AT2019dsg and AT2019fdr, respectively. All panels are shown for model  M-IR  in the observer's frame.
}
\label{fig:MIR_spectra}
\end{figure*}

%%%%%%%%%%%%%% Section 3: Neutrino and EM cascade emissions %%%%%%%%%%%%%
\section{Neutrino and EM cascade emissions}
\label{sec:cascade}

In this section, we present the spectral energy distributions (SEDs), i.e., $E_i^2 \mathcal F_i=t_{\rm fs}^{-1}E_i^2dN_i/dE_i/(4\pi d_L^2)$, of neutrino and electromagnetic cascade emissions for M-IR (Sec.~\ref{subsec:spec_IR}) and M-OUV (Sec.~\ref{subsec:spec_OUV}) scenarios of AT2019dsg and AT2019fdr, where $d_L$ is the luminosity distance to the TDE and the red shift corrections to fluxes and energies are not explicitly shown in this equation for simplicity. To establish a connection between our models and observations, we provide also light curves in the relevant X-ray and gamma-ray energy bands, and compare our predictions with measured data points or upper limits.

\subsection{M-IR: extended radiation zone ($R\simeq R_{\rm IR}$)}
\label{subsec:spec_IR}

Using the methodology described in Section \ref{sec:TDE} and the parameter values for M-IR scenario provided in Tab. \ref{tab:params}, we present in Fig. \ref{fig:MIR_spectra} the SEDs and light curves for AT2019dsg (upper row) and AT2019fdr (lower row) in the observer's frame. The left column  shows the SEDs at neutrino detection time $t_\nu$, depicting the thermal IR, OUV and X-ray spectra (in blue dotted curves denoted as `BB'), as well as the single-flavor neutrino (e.g., $\nu_\mu$, in magenta curves) and EM cascade spectra spanning from MeV to 100 PeV. The black curves represent the cascade spectra without taking into account the $\gamma\gamma$ attenuation with extragalactic background light (EBL) and cosmic microwave background (CMB). The different components of the electromagnetic cascades, including `$pp/p\gamma-$SY/IC', `$\gamma\gamma-$SY/IC', `BH-SY/IC', and `$\pi^0$ decay', are depicted as the blue dashed, orange dash-dotted, green dotted, and red solid curves, respectively. The gray shaded areas in the figure highlight the energy ranges of cosmic $\gamma\gamma$ attenuation with EBL and CMB. One notable feature of the EM cascade spectra is the pronounced dips in the energy range $\sim1~\rm GeV-1~PeV$. These dips can be attributed to the in-source $\gamma\gamma$ absorption with thermal photons. Considering the characteristic energy of thermal photons $E_{\rm BB}$, the $\gamma\gamma$ attenuation occurs at $E_{\gamma\gamma}\sim(m_ec^2)^2/E_{\rm BB}\simeq2.6{~\rm GeV~}(E_{\rm BB}/100~{\rm eV})^{-1}$ in the laboratory frame. For instance, X-ray photons with energy of $k_BT_X=72$ eV will lead to the first absorption dip at $\sim4$ GeV in the black curve of the upper left panel.

For AT2019dsg, the blue, red, and green shaded areas in the upper left panel illustrate the energy ranges for the observations or limits using XRT/XMM/NICER (0.3-10 keV), \emph{Fermi} LAT (0.1-800 GeV) and High-Altitude Water Cherenkov Observatory (HAWC, 0.3-100 TeV) \citep[see][and references therein]{2019GCN.25936....1A,2021NatAs...5..510S,2021MNRAS.504..792C,2022PhRvL.128v1101R,2021A&A...647A...1P}. The corresponding light curves are shown in the upper right panel. The orange and green data points are the energy-integrated upper limits obtained by \emph{Fermi} and HAWC, whereas the XRT/XMM/NICER observations are plotted as blue points. The \emph{Fermi} $\gamma$-ray up limits are obtained in three time slots, $t-t_{\rm pk}\sim-20-100{~\rm d},~{100-200~\rm d}$ and $200-500~\rm d$. The upper limit in the first time slot is more stringent according to the upper right panel of Fig. \ref{fig:MIR_spectra} and will be used in Sec. \ref{sec:constraints} for the constraints on the $R-E_{p,\rm max}$ plane. The multiwavelength light curves indicate that the $\gamma$-ray upper limits are satisfied in the M-IR scenario, although the limits are not far way from the predictions. This also implies that much higher predicted neutrino fluxes would contradict the $\gamma$-ray limits.

For reference purposes, we include the thermal OUV light curve, e.g., $L_{\rm OUV}/4\pi d_L^2$, as the black dashed curve. The peaks of EM cascade light curves appear roughly $30-50$ days after the OUV peak, which is commonly referred to as the time delay. In the M-IR scenario, where the radiation zone is calorimetric but optically thin to $p\gamma$ interactions ($\tau_{p\gamma}^{\rm fs}<1$), the time delay observed in the EM cascade emissions, primarily initiated via $p\gamma$ processes, can be interpreted as the $p\gamma$ time scale, e.g., $t_{p\gamma}\sim30-50$ days\footnote{{Typically, one should take into account the rates of $\gamma\gamma$ attenuation, pion/muon decay, and secondary electron synchrotron/inverse Compton to estimate cascade time scale. In this calculation, the interactions mentioned here are much faster than $p\gamma$ process and do not contribute significantly to the time delay. We show the interaction rates for secondary processes in Appendix \ref{app:rates} to support this statement.}}. This reflects the time-dependent nature of EM cascade as a redistribution of energy via a series of successive radiation processes. Hence, if future observed TDEs exhibit time delays in their X-ray and $\gamma$-ray light curves, one can employ $p\gamma$ optically-thin models to interpret and understand such temporal signatures. {Furthermore, we observe that the orange light curve is additionally delayed compared to the X-ray and VHE $\gamma$-ray light curves. In the energy range of 0.1 - 800 GeV, $\gamma\gamma$ attenuation is predominantly influenced by OUV as well as X-ray photons. Close to the peak time of the OUV blackbody light curve, where the OUV photon density is higher, the resulting broadband integrated $\gamma$-ray flux is more heavily attenuated. This could lead to the later emergence of the $\gamma$-ray peak.}

The rapid (exponential) decay of observed X-ray light curve in the time interval $0\lesssim t-t_{\rm pk}\lesssim 50~\rm d$ (illustrated as the blue dashed line in the upper right panel of Fig. \ref{fig:MIR_spectra}) might be caused by the cooling of the accretion disk or by dust obscuration. Interestingly, there is an additional  X-ray (Swift-XRT) data point around $t-t_{\rm pk}=100$ d which was identified in \citet{2021MNRAS.504..792C} and does not fit this rapid decay picture. Notably, the X-ray cascade emission can describe that data point, which means that Swift may have actually seen an electromagnetic cascade signature there. One may speculate that if the populations of neutrino-emitting TDEs and X-ray bright TDEs have overlap, the often complicated and puzzling behavior of X-ray observations may in some cases be explained by the contribution of the EM cascade, which can be substantially delayed with respect to the BB peak (see e.g. example for AT2019fdr below). The identification of a spectral hardening  in the X-ray range (see upper left panel) may provide future evidence.

Our calculation also provide the neutrino spectra and light curves, as shown in the magenta curves in Fig. \ref{fig:MIR_spectra}. Since $p\gamma$ interactions dominate the neutrino production compared to the $pp$ process, we show only the single-flavor neutrino spectra obtained from the $p\gamma$ channel in the left column. The black arrows indicate the most-likely energies of the coincident neutrinos. {From the upper right panel, we observe that the energy-integrated neutrino flux, $L_\nu/4\pi d_L^2$, exhibits a similar time delay compared to the X-ray light curve. The reason is that, in the M-IR scenario of AT2019dsg, despite the radiation zone being nearly calorimetric to UHECRs, i.e., $\tau_{p\gamma}^{\rm cal}=t_{p\gamma}^{-1}/t_{p,\rm esc}^{-1}\sim1$, the $p\gamma$ interaction rate is lower than the free streaming rate (see the left panel of Fig. \ref{fig:timescales}). This implies that the $p\gamma$ interactions are efficient (over the escape timescale) but not very fast} \citep[see][for a more detailed discussion]{2023ApJ...948...42W}. %Therefore, the neutrino luminosity $L_\nu=t_{\rm fs}^{-1}\int E_\nu (dN_\nu/dE_\nu)dE_\nu$ is proportional to the in-source proton luminosity $L_p$, as they are linked via
%\begin{equation}
%   \left.E_\nu^2\frac{dN_\nu}{dE_\nu}\sim \frac{1}{8}\min[1,\tau_{p\gamma}^{\rm cal}]E_p^2\frac{dN_p}{dE_p}\right\vert_{E_p\simeq20E_\nu}.
%   \label{eq:neutrinos}
%\end{equation}
%Combining this relation with equation \ref{eq:CR_luminosity}, we have $L_\nu\propto L_{\rm OUV}$. We also observe that the neutrino fluxes at the energies of the coincident neutrinos (indicated by the black arrows in the left column) are relatively low. This is due to the fact that the peak energies of the predicted neutrino spectra, given by $E_{\nu,\rm pk}\simeq E_{p\gamma,\rm max}/20\simeq 50~{\rm PeV}(E_{p\gamma,\rm max}/10^9~\rm GeV)$, are much higher than the measured energies. In the equation, $E_{p\gamma,\rm max}\simeq 10^9$ GeV represents the proton energy at which the interaction rate of $p\gamma$ interactions, $t_{p\gamma}^{-1}$, reaches its peak value.

\begin{figure*}[tp]\centering
\includegraphics[width=0.49\textwidth]{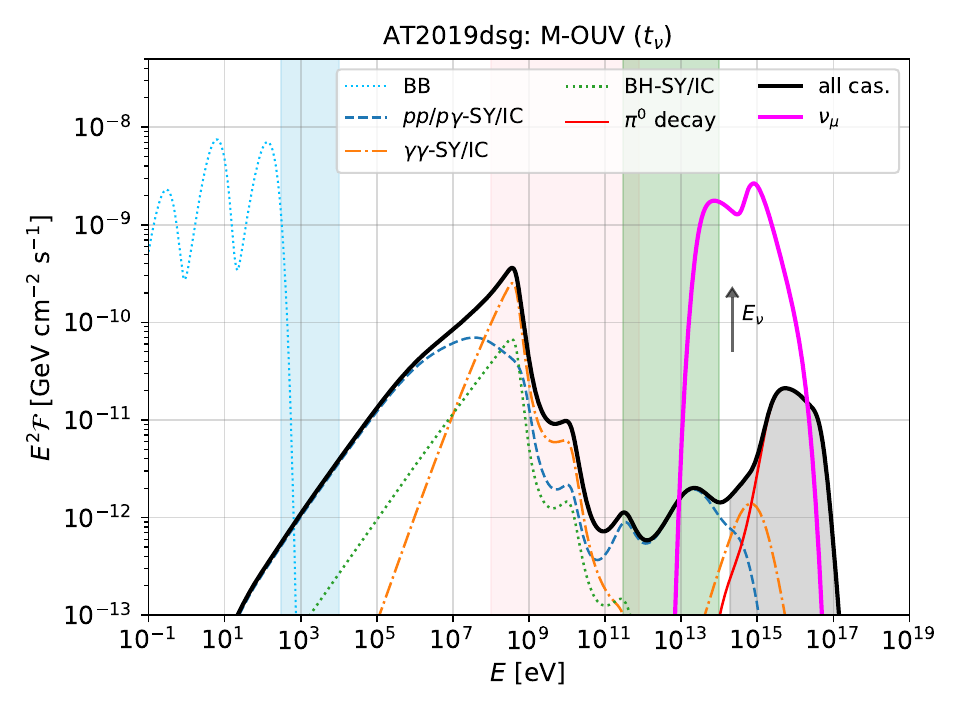}
\includegraphics[width=0.49\textwidth]{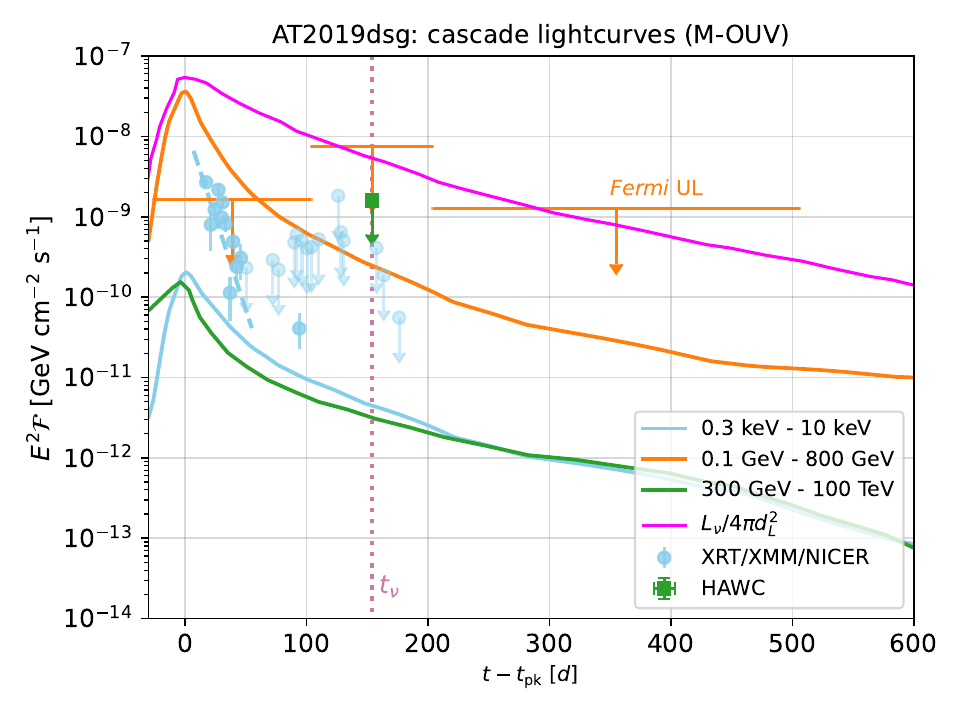}
\includegraphics[width=0.49\textwidth]{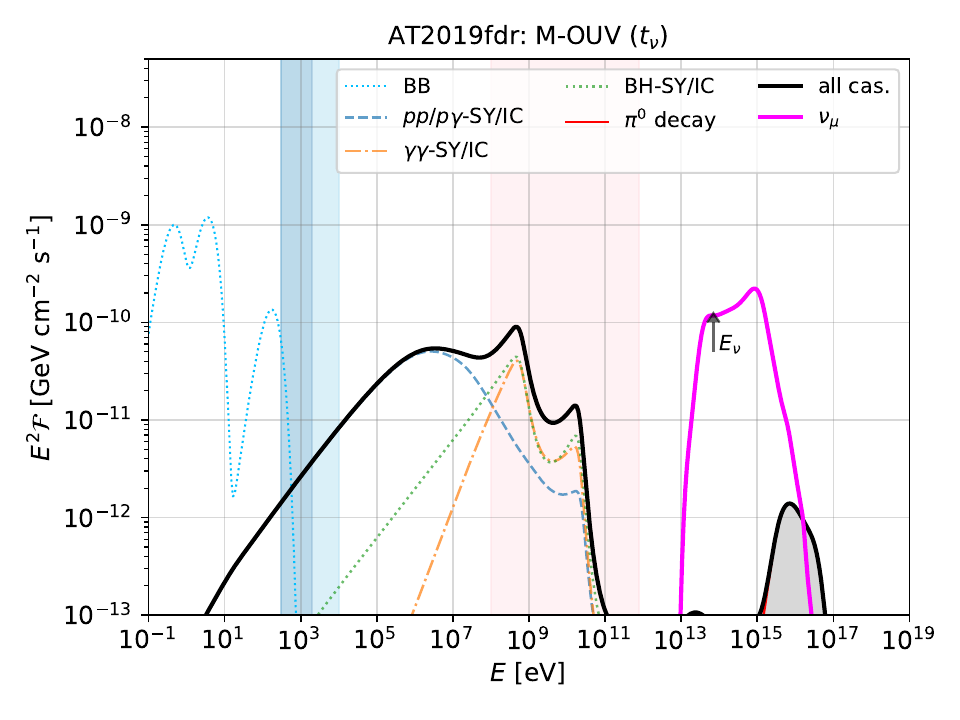}
\includegraphics[width=0.49\textwidth]{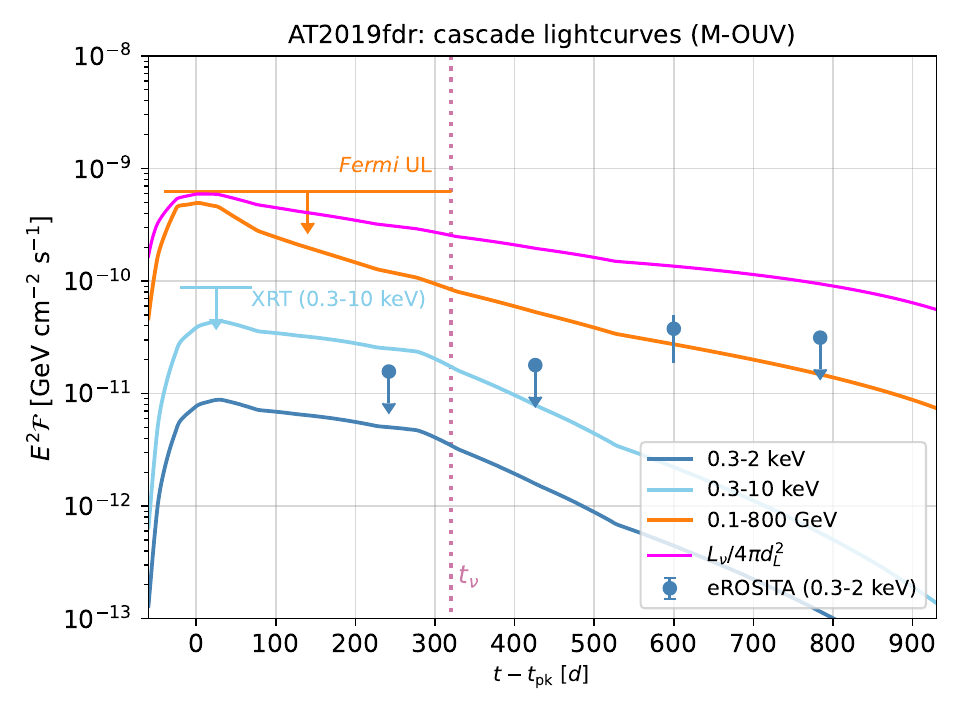}
\caption{Electromagnetic cascade spectra and neutrino spectra at $t_\nu$ (left column), and neutrino, $\gamma$-ray and X-ray light curves (right column), for the M-OUV model. The curves, areas, and data points have the same meaning as in Fig. \ref{fig:MIR_spectra}.}
\label{fig:MOUV_spectra}
\end{figure*}

Correspondingly, the results for AT2019fdr are displayed in the lower row of Fig. \ref{fig:MIR_spectra}. The areas, data points, and curves have the same interpretation as in the AT2019dsg cases. However, in this case, the deep blue areas, data points, and curves specifically represent the observations and predictions in the energy range of 0.3$-$2 keV, associated with the eROSITA upper limits. 
The upper limit on $\gamma$-ray emissions is only measured within the time interval $-20~\rm{d}\lesssim t-t_{\rm pk}\lesssim300~\rm{d}$. Additionally, X-ray follow-up observations using Swift XRT provide an upper bound around the time of the OUV peak, while the later time constraints (represented by the green points) are measured by eROSITA. The M-IR scenario of AT2019fdr is consistent with the X-ray and $\gamma$-ray observational limits. However, it is challenging for the EM cascade models to explain the late-time X-ray peak observed at $t-t_{\rm pk}\simeq 600$ days by eROSITA. This particular signature may be produced by other mechanisms or regions that are distinct from the neutrino-emitting zones discussed here, especially considering that it appears roughly 300 days after the detection of neutrinos. 

{Time delays in the X-ray and $\gamma$-ray light curves are also observed, and can be attributed jointly to the $p\gamma$ interaction times and the peak time of the IR light curve.} Since a larger radius of the radiation zone, e.g., $R=2.5\times10^{17}$ cm, is used for AT2019fdr, compared to that of AT2019dsg ($R=5\times10^{16}$ cm), a prolonged time delay of approximately 100$-$200 days is expected due to the decreasing $p\gamma$ interaction rate\footnote{{The AT2019fdr timescales are shwon in Figure 9 of \cite{2023ApJ...948...42W}.}}. {Moreover, the IR light curve of AT2019fdr exhibits a strong peak at around 300 days, compared to the relatively flat IR light curve of AT2019dsg. This peak in the photon distribution, combined with the $t_{p\gamma}$ value, contributes to the time delays observed in X-ray and $\gamma$-ray emissions.} Unlike the case of AT2019dsg, the neutrino light curve of AT2019fdr no longer simply follows the time dependence of OUV luminosity. The reason is that the neutrino production is determined not only by the $p\gamma$ interaction rate but also the in-source proton distributions. For AT2019fdr, the protons are more strongly confined within the radiation zone in comparison to AT2019dsg. As a consequence, there is a buildup or ``piling-up" of protons, which enhances the late-time neutrino productions and lead to the time delay\footnote{Here, we only provide a qualitative explanation for the time dependence of neutrino emission, as this paper primarily focuses on the EM cascades. A more quantitative discussion for the neutrino light curves of AT2019dsg and AT2019fdr can be found in \cite{2023ApJ...948...42W}.}.

In summary, in the M-IR scenario where the radiation zone can extend to the dust radius, the $p\gamma$ optical depth is typically optically thin. This leads to time delays in the EM cascade light curves. One advantage of this scenario is that the predicted X-ray and $\gamma$-ray emissions obey well with the observational upper limits, as shown in the right column of Fig. \ref{fig:MIR_spectra}. Remarkably, the M-IR scenario of AT2019dsg is capable of explaining the X-ray observations during the time interval $50~{\rm d}\lesssim t-t_{\rm pk}\lesssim 100~\rm d$.

\subsection{M-OUV: compact radiation zone ($R\ll R_{\rm IR}$)}\label{subsec:spec_OUV}

Here we consider more compact radiation zones in the M-OUV scenarios and discuss the EM cascade signatures in the $p\gamma$ optically thick cases, for AT2019dsg and AT2019fdr. Similar to the Fig. \ref{fig:MIR_spectra}, the SEDs and light curves for the M-OUV scenarios are demonstrated in Fig. \ref{fig:MOUV_spectra}. The parameters used here are summarized in Tab. \ref{tab:params}. We find that the neutrino peak fluxes in the M-OUV scenarios for both AT2019dsg and AT2019fdr are significantly higher compared to the M-IR scenarios. This enhancement is attributed to the efficient $p\gamma$ interactions in the M-OUV scenarios, as indicated in the right panel of Fig. \ref{fig:timescales}, facilitated by the much denser target photons. As a result, the neutrino fluxes in the M-OUV scenarios are enhanced by a factor of $2-3$ compared to those in the M-IR scenarios. Additionally, the peaks of the neutrino spectra closely align with the energies of the coincident neutrinos. This, in combination with the high rate of efficient $p\gamma$ interactions, makes TDEs in M-OUV scenarios as the promising neutrino emitters. From the right column, we find that the neutrino light curves exhibit the same shape as the OUV luminosities ($L_{\rm OUV}/4\pi d_L^2$ is not shown for simplicity). This can be straightforwardly interpreted by the optically thick nature of the radiation zones, i.e., the protons interact very quickly before they can escape even compared to the size of the region, where the condition $\tau_{p\gamma}^{\rm fs}=t_{p\gamma}^{-1}/t_{p,\rm fs}^{-1}>1$ is satisfied, as depicted in the right panel of Fig. \ref{fig:timescales}. Consequently, no significant time delays are observed.

The EM cascade fluxes, similar to the neutrinos, are enhanced as well, particularly at the peak time $t_{\rm pk}$, and there are no significant time delays with respect to the OUV curve.  For AT2019dsg, the upper right panel of Fig. \ref{fig:MOUV_spectra} demonstrates that the \emph{Fermi} $\gamma$-ray upper limits are violated in the M-OUV scenarios with the radius of $R=5\times10^{14}~\rm cm$. It is worth noting that when the radius is increased to $10^{15}-10^{16}~\rm cm$, the peak flux shows a significant drop, and the upper limits from \emph{Fermi} observations are no longer violated. This is observed in the M-OUV scenario of AT2019fdr, as shown in the lower right panel. In this discussion, we provide a intuitive conclusion regarding the \emph{Fermi} constraints. A more detailed analysis of the implications of these limits on the $R-E_{p,\rm max}$ plane will be presented in Sec. \ref{sec:constraints}.

One common feature of the EM cascade SEDs in the M-IR and M-OUV scenarios is that the `$pp/p\gamma-$SY/IC' (blue dashed curves)  and `$\gamma\gamma-$SY/IC' (orange dashed-dotted curves) components in the left columns of Figs. \ref{fig:MIR_spectra} and \ref{fig:MOUV_spectra} give rise to the low-energy ($\sim0.1-10$ MeV) and high-energy ($\sim 0.1-10$ GeV) humps, respectively. Noting that in $p\gamma$ channel around $5\%$ of the proton energy is inherited by secondary leptons, we estimate the characteristic Lorentz factor of electrons and positrons to be $\gamma_{e,p\gamma}\simeq (E_{p,\rm cal}/20)/(m_ec^2)\sim 10^8 (E_{p,\rm cal}/(10^6 \, \mathrm{GeV}))$, where $E_{p,\rm cal}\simeq10^5-10^6$ GeV is the proton energy at which the radiation zone becomes calorimetric, e.g., $t_{p\gamma}^{-1}\sim t_{\rm esc}^{-1}$. We then infer the peak energy of synchrotron radiation from $p\gamma$ electrons and positrons
\begin{equation}
E_{p\gamma,\rm sy}=\frac{3}{4\pi}h\gamma_{e,p\gamma}^2\frac{eB}{m_ec}\simeq 17{~\rm MeV}\left(\frac{E_{p,\rm cal}}{10^6~\rm GeV}\right)^2,
\label{eqn:synchrotron_energy}
\end{equation}
where the magnetic field strength $B=0.1$ G is used. The corresponding inverse Compton components are typically invisible due to $\gamma\gamma$ absorption. Regarding the `$\gamma\gamma-$SY/IC' channel, the interpretation is more complicated since the $\gamma\gamma$ cross section exhibits slow decay beyond the threshold, suggesting a wide energy range of absorption. From the EM cascade spectra, we observe that the strongest absorption occurs close to the peak energies of $\gamma$-rays from $\pi_0$ decays, as predicted by $E_{\gamma\gamma,\rm max}\sim E_{p,\rm th}/10$, where $E_{p,\rm th}$ is defined in Eq. (\ref{equ:pgamma}) and we consider that approximately 10\% of the proton energy is transferred to $\gamma$-rays via the $\pi_0\rightarrow\gamma\gamma$ channel. The peak energy of the resulting synchrotron emission, produced by electrons of $E_{\gamma\gamma,\rm max}/2$, is given by $E_{\gamma\gamma,\rm sy}\sim\mathcal O(1{~\rm TeV})(E_\gamma/0.1\rm eV)^{-2}$, where $E_\gamma$ is the energy of target photons. In this case, the synchrotron emission in the `$\gamma\gamma-$SY/IC' channel suffers from attenuation before reaching its maximum $E_{\gamma\gamma,\rm sy}$, which explains the sharp spikes appearing in the EM cascade spectra. This semi-theoretical estimation justifies the numerical results shown in Figs. \ref{fig:MIR_spectra} and \ref{fig:MOUV_spectra}.

\begin{figure*}[htp]\centering
\includegraphics[width=0.49\textwidth]{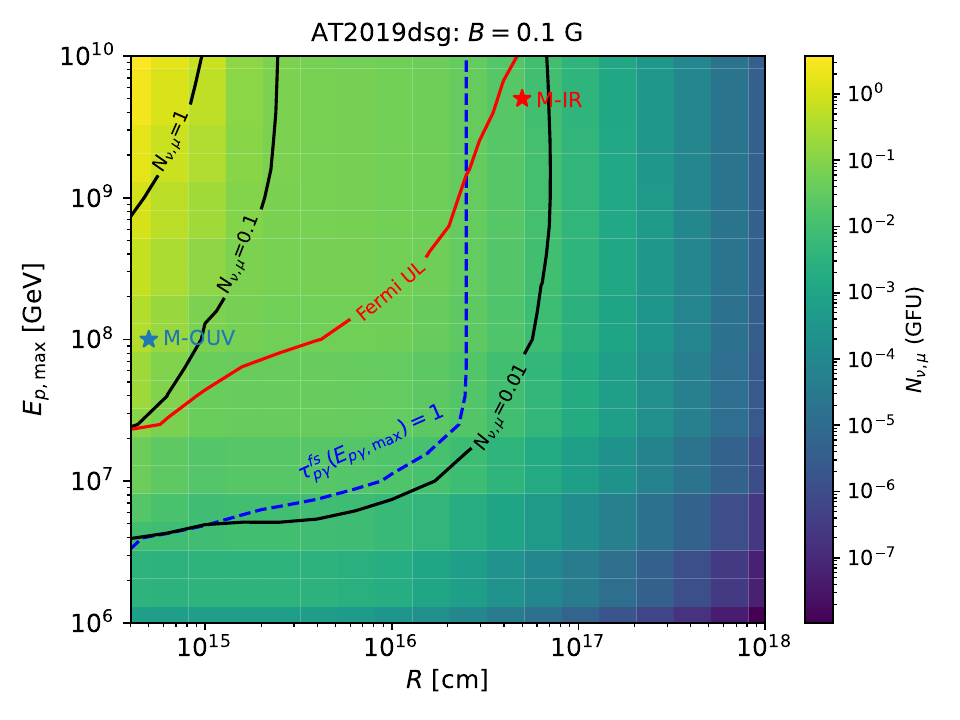}
\includegraphics[width=0.49\textwidth]{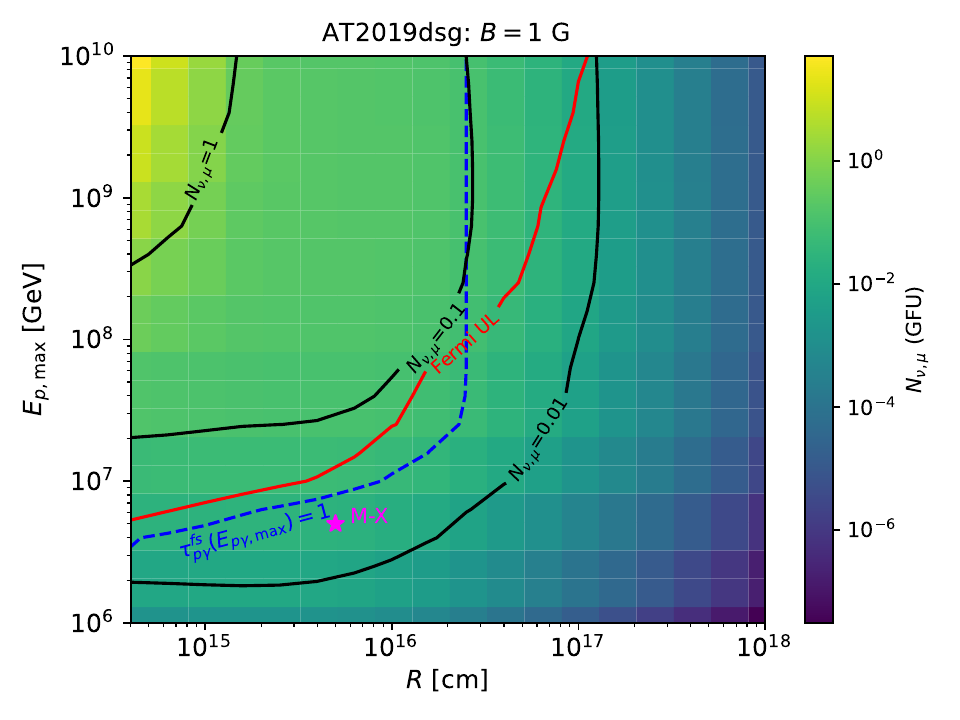}
\includegraphics[width=0.49\textwidth]{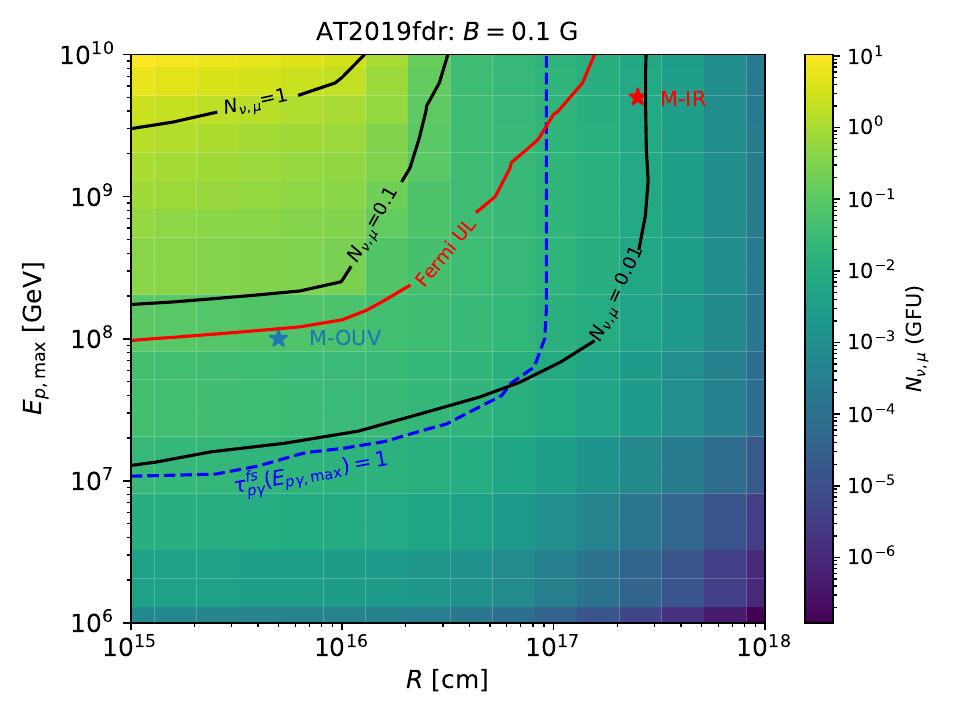}
\includegraphics[width=0.49\textwidth]{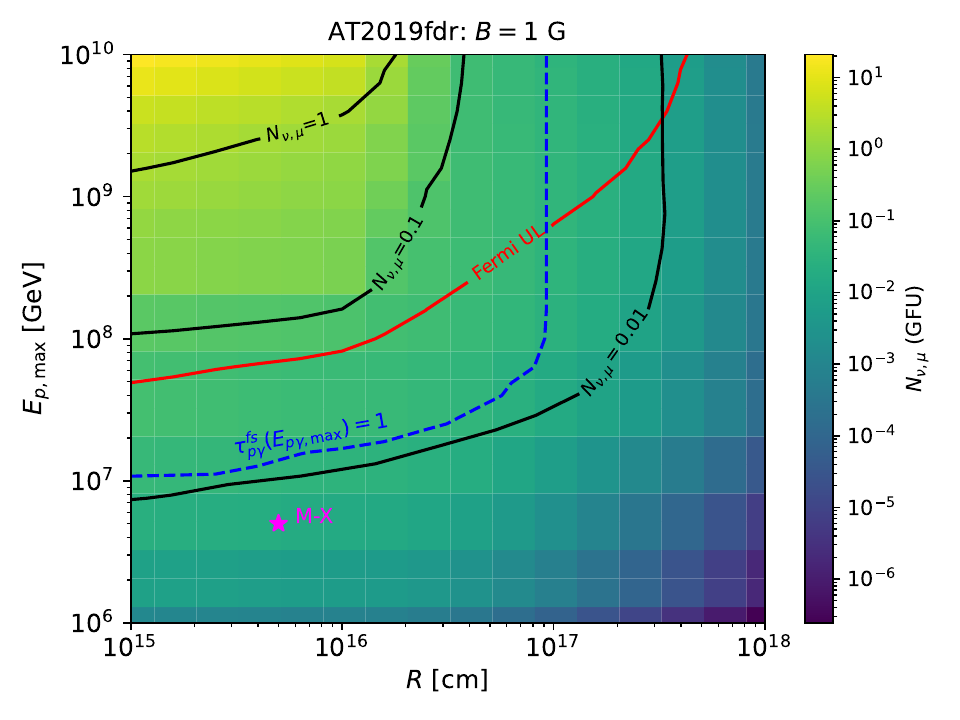}
\caption{Number of expected single-flavor neutrino events as a function of  $R$ and $E_{p,\rm max}$ as given by the color scales; we use the IceCube Gamma-Ray Follow-up (GFU) effective areas to compute the neutrino numbers.  The panels correspond to AT2019dsg (upper row) and AT2019fdr (lower row), and  magnetic field strengths $B=0.1$ G (left column) and $1$ G are (right column). The \emph{Fermi} $\gamma$-ray upper limits on the $R-E_{p,\rm max}$ plane are shown as red curves (the parameters in the upper left corners are excluded), and the regions where the optical thickness  $\tau_{p\gamma}^{\rm fs}(E_{p\gamma,\rm max})=1$ are shown as blue dashed curves (the parameters in the upper left corners are optically thick); here the $p\gamma$ optical depth is evaluated at $t_\nu$ and $E_{p\gamma,\rm max}$ and the definition of $E_{p\gamma,\rm max}$ can be found in the main text. Moreover, the models M-X, M-IR and M-OUV models are marked by stars. }
\label{fig:constraints}
\end{figure*}

%%%%%%%%%%%%%% Section 4: Constraints %%%%%%%%%%%%%
\section{$\gamma$-ray constraints on $E_{p,\rm max}$ and $R$}
\label{sec:constraints}

In the previous sections, we examined EM cascade and neutrino emissions in two special scenarios, M-IR and M-OUV, for AT2019dsg and AT2019fdr. It is useful to investigate the model-predicted neutrino numbers as a function of $R$ and $E_{p,\rm max}$ and explore the potential constraints on source parameters using \emph{Fermi} $\gamma$-ray up limits.

We choose $R\in [5\times10^{14}~{\rm cm}, 10^{18}~{\rm cm}]$ ($[10^{15}~{\rm cm}, 10^{18}~{\rm cm}]$ for AT2019fdr) and $E_{p,\rm max}\in[10^6~{\rm GeV}, 10^{10}~{\rm GeV}]$ as the fiducial ranges to allow for a continuous interpolation among the different models. From the left panels of Figs. \ref{fig:MIR_spectra} and \ref{fig:MOUV_spectra}, we notice that the \emph{Fermi} $\gamma$-ray constraints are more stringent compared to the X-ray and HAWC $\gamma$-ray observations. Hence, we use the \emph{Fermi} constraints in the energy range $0.1-800$ GeV to study the permissible parameter sets that are compatible with observations. 

Our main results for parameter constraints are illustrated in Fig. \ref{fig:constraints}. The upper and lower rows correspond to the TDEs AT2019dsg and AT2019fdr, whereas the magnetic fields of $B=0.1$ G and $1.0$ G are used in the left and right columns respectively to demonstrate the impact of different magnetic field strengths. The M-OUV and M-IR scenarios are indicated by blue and red stars, respectively. We also include the parameters (e.g., $R=5\times10^{15}{~\rm cm},~ E_{p,\rm max}=5\times10^{6}{~\rm GeV}$ and $B=1.0$ G) for the M-X scenarios discussed in \citet{2023ApJ...948...42W} as the magenta stars. The red contours represent the \emph{Fermi} upper limits (denoted as `\emph{Fermi} UL'). For a TDE radiation zone described by the parameter sets above the red contours, the EM cascade emission will overshoot the \emph{Fermi} upper limits. This conclusion is consistent with the results in Sec. \ref{sec:cascade}, which show that the compact radiation zones violate the upper limits. 

To understand the compactness of the radiation zone, we present the critical conditions for the source to be $p\gamma$ optically thick in the dashed blue contours as well. These conditions are obtained through $\tau_{p\gamma}^{\rm fs}(E_{p\gamma,\rm max})=1$, where $E_{p\gamma,\rm max}<E_{p,\rm max}$ is the proton energy at which $t_{p\gamma}^{-1}$ reaches its maximum. We find that there exists a critical radius beyond which the system can no longer be $p\gamma$ optically thick even for UHE protons. The reason is that $t_{p\gamma}^{-1}$ is limited by the maximum target photon density and the optical depth decreases with respect to $R$, e.g., $\tau_{p\gamma}^{\rm fs}=t_{p\gamma}^{-1}/t_{\rm fs}^{-1}\propto R^{-1}$, as $t_{p\gamma}^{-1}\propto R^{-2}$ and $t_{\rm fs}^{-1}\propto R^{-1}$. Above the blue dashed lines, the EM cascade emissions do not exhibit significant time delays due to the faster secondary radiation processes compared to photon free streaming. However, below the blue dashed line, a time delay on the scale of $t_{p\gamma}$ can be expected. 

To discuss the implications of \emph{Fermi} $\gamma$-ray constraints on neutrino detection, we show the model-predicted neutrino numbers as a function of $R$ and $E_{p,\rm max}$ in the form of meshed color maps with the red contours in Fig. \ref{fig:constraints}. The IceCube gamma-ray follow-up (GFU)\footnote{The GFU neutrino numbers, estimated using IceCube GFU effective areas \citep{2016JInst..1111009I}, are more suitable when comparing the model predictions to actual follow-up observations, whereas point source (PS) neutrino numbers are typically used for independent point-source analyses.} neutrino numbers can be estimated via
\begin{equation}
    N_{\nu,\mu}({\rm GFU})=\int dE_\nu A_{\rm eff}(E_\nu)F_\nu(E_\nu,t_\nu)
\end{equation}
where $F_\nu(E_\nu,t_\nu)=\int^{t_{\nu}}dt \mathcal F_\nu(E_\nu,t)$ is the cumulative neutrino fluence until $t_\nu$ and $A_{\rm eff}$ is the IceCube GFU effective area. %The maximum and minimum neutrino energies in the integral are determined by the 99.73\% confidence level expected neutrino energy ranges, e.g., $E_{\nu,\rm min}\simeq6.0\times10^4~{\rm GeV}~(3.3\times10^{4}~\rm GeV)$ and $E_{\nu,\rm max}=3.5\times10^6~{\rm GeV}~(1.0\times10^{6}~\rm GeV)$ for At2019dsg (AT2019fdr) \cite[see][for the detailed descriptions]{2023ApJ...948...42W,2018A&A...615A.168P}. 
The black contours in Fig. \ref{fig:constraints} indicate the single-flavor neutrino numbers $N_{\nu,\mu}$(GFU)$=1,~0.1$, and 0.01. From this figure, we find that to avoid overshooting the \emph{Fermi} $\gamma$-ray up limits, the model-predicted neutrino numbers detected by IceCube are constrained to be $N_{\nu,\mu}({\rm GFU})\lesssim 0.1$. This conclusion applies to both cases with $B=0.1$ G (left column) and 1.0 G (right column).
Specifically, our findings indicate that the M-X scenarios share similarities with the M-IR cases. Both scenarios are situated in the $p\gamma$ optically thin regime, and they both predict a similar neutrino event rate and satisfy the upper limits set by the \emph{Fermi} observations.

When comparing the left column with the right column, it is worth noting that a stronger magnetic field would result in a higher predicted neutrino number, assuming fixed source parameters. Additionally, a stronger magnetic field would also increase the likelihood of exceeding the $\gamma$-ray limits due to enhanced confinement of charged particles within the radiation zones. This enhanced confinement would lead to more efficient $p\gamma$ interactions and subsequent EM cascades.

From a multi-messenger perspective, we observe a correlation among three regions: high optical thickness for $p\gamma$ interactions, high rates of neutrino events, and high fluxes of $\gamma$-rays.  Specifically, high neutrino event rates indicate the presence of intense EM cascades, which are facilitated by efficient and rapid $p\gamma$ interactions. The fact that time delays in neutrino signals with respect to OUV peaks are observed while no $\gamma$-ray induced by EM cascades has been seen implies that the neutrino production cannot be too efficient, e.g., the predicted upper limit for neutrino event rates from \emph{Fermi} constraints ranges between 0.01 and 0.1 events per TDE, as one can read off from the Fig. \ref{fig:constraints}, and the system is probably optically thin for $p\gamma$ interactions. Future multi-messenger observations, in particularly incorporating the $\gamma$-rays measurements, could shed more lights onto the radiation processes and physical conditions in the radiation zones.

%%%%%%%%%%%%%% Section 5: Discussion %%%%%%%%%%%%%
\section{Discussion}\label{sec:discussion}

\subsection{EM cascades from TDEs}\label{subsec:discussion_cascade}

As noted before in Sec. \ref{sec:cascade}, the EM cascade emissions exhibit some commonalities across different scenarios and TDEs. One prominent feature of the SEDs is the deep dip in the energy range approximately from 10-100 GeV to 1-10 PeV, which is a result of the in-source $\gamma\gamma$ attenuation with thermal IR, OUV, and X-ray photons. Therefore, only a part of the whole EM cascade spectra, ranging from approximately 1-10 keV to 1-10 GeV, can be measured by eROSITA, Swift XRT, XMM-Newton, NICER, and \emph{Fermi} LAT in the observer's frame. Additionally, the multi-wavelength light curves of the EM cascades shown in the right columns of Figs. \ref{fig:MIR_spectra} and \ref{fig:MOUV_spectra} also display unique signatures with respect to the IR, OUV and X-ray BB emissions. {For instance, in M-IR cases where the $p\gamma$ interactions are slow, the peaks of EM cascade light curves usually emerge after the OUV peak as well, leading to the time delays.} More generally, similar time delays are also expected if the radiation zone has a large radius or the maximum energy of injected protons is low, as indicated by the regions below the blue dashed lines in Fig. \ref{fig:constraints}. In this case, X-ray and $\gamma$-ray follow-ups after the identification of OUV peaks would reveal these delayed emissions. 
The fully time-dependent treatment in this paper provides a more accurate modeling for the neutrino and EM cascade emissions compared to steady-state approximations, particularly in cases where the interactions are efficient (calorimetric) but not sufficiently fast ($p\gamma$ optically thin). On the other hand, if the radiation zone is compact and $E_{p,\rm max}$ is high (see the regions above the blue dashed lines in Fig. \ref{fig:constraints}), the system can be $p\gamma$ optically thick and the light curves of EM cascade emissions reach their peak roughly at the same time with the neutrino and thermal OUV light curves. In both cases, the X-ray and $\gamma$-ray fluxes can reach the detectable levels, making the follow-up observations possible. %However, the absence of $\gamma$-ray measurements of TDEs detected so far implies that the radiation zones are probably $p\gamma$ optically thin for most TDEs, but some of them, such as AT2019dsg and AT2019fdr, can be simultaneously calormetric to produce 0.01-0.1 neutrino events. This prediction is consistent with the expected neutrino numbers from IceCube GFU searches, as listed in Tab. \ref{tab:params}.

In addition to the neutrino-coincident TDEs AT2019dsg and AT2019fdr, whose neutrino counterparts are identified via follow-up searches, AT2019aalc stands out with the highest dust echo flux among all ZTF transients and is also likely to be associated with the neutrino event IC 191119A. The SMBH mass and AT2019aalc is comparable to that of AT2019fdr, while the \emph{Fermi} limits and the time difference of neutrino arrival measured relative to the OUV peak are similar to those of AT2019dsg. However, no significant X-ray limits or observations for AT2019aalc has been reported. \cite{2023ApJ...948...42W} systematically studied the neutrino emissions from all three neutrino-coincident TDEs and confirmed the similarities between them. Despite not shown explicitly in this paper, we investigated the EM cascade emissions from AT2019aalc as well and got similar conclusions with AT2019dsg and AT2019fdr. For example, high neutrino event rates imply strong EM cascades and violation of \emph{Fermi} upper limits. For instance, the M-OUV scenario of AT2019aalc would also overshoot the \emph{Fermi} upper limits obtained by 207 days of observations from the discovery of the optical emission. This on the other hand demonstrates that our model can be widely applied to TDEs.

\begin{figure}
    \centering
    \includegraphics[width=0.5\textwidth]{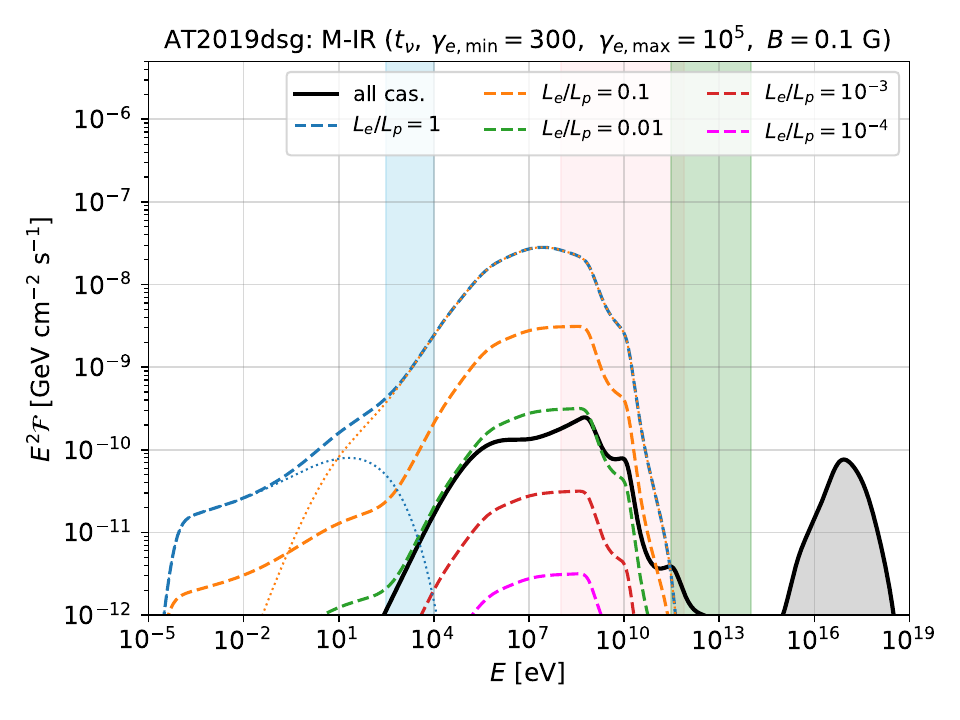}
    \caption{Comparison of EM cascade emission (black solid curve) with the leptonic contribution (dashed curves). The shaded areas have the same meaning as those in the upper left panel of Fig. \ref{fig:MIR_spectra}. To obtain the EM cascade spectrum, the parameters used in M-IR scenario of AT2019dsg are applied (e.g., $B=0.1$ G). From the upper to lower dashed curves, the ratio of injected electron luminosity to the proton luminosity, $K_{e/p}\equiv L_e/L_p$, varies from 1 to $10^{-4}$. The blue and orange dotted curves show the synchrotron and inverse Compton components of the $L_e/L_p=1$ case.} 
    \label{fig:leptonics}
\end{figure}

\subsection{Leptonic loading in TDEs}\label{subsec:discussion_leptonic}

The EM cascade emissions discussed in this work are initiated by hadronic processes without taking into account the injection of primary electrons or positrons. In practice, we also expect the acceleration of leptons at the acceleration sites where protons are energized. However, the accretion power converted to leptons is typically less than that of protons. The leptonic loading factor ($K_{e/p}$), defined as the ratio of the luminosities of injected electrons/positrons to protons, is typically less than unity, e.g., $K_{e/p}\equiv L_e/L_p<1$. Simulations of non-relativistic diffusive shock acceleration in supernova remnants have demonstrated that the efficiency of electron energy deposition, compared to proton acceleration, is at least two orders of magnitude lower, e.g., $K_{e/p}\lesssim10^{-2}$ \citep[see, e.g.,][]{2011JApA...32..427J,2012A&A...538A..81M}. Furthermore, the particle-in-cell simulation gives a similar value, e.g., $K_{e/p}\simeq10^{-3}$ \citep{2008JCAP...01..018K,2014ApJ...783...91C,2015PhRvL.114h5003P}. For relativistic shocks, a leptonic loading factor on the order of $K_{e/p}\sim\mathcal O(10^{-4}-1)$ is widely used in active galactic nucleus (AGN) models \citep[roughly corresponding to the baryon loading factor $\xi_{\rm cr}\sim1-10^4$, see e.g.,][]{2014PhRvD..90b3007M,2014ApJS..215....5K,2019NatAs...3...88G,2020ApJ...891..115P}. Therefore, in order to obtain a more comprehensive understanding of the radiation zone, we compare the EM cascade spectra with the synchrotron and inverse Compton emissions (in the synchrotron self-Compton picture) from leptonic loading by varying $K_{e/p}$ from $10^{-4}$ to the extreme value $K_{e/p}=1.0$. We consider a power law electron/positron injection $Q_e\propto\gamma_e^{-2}\exp(-\gamma_e/\gamma_{e,\rm max})$ with the minimum and maximum Lorentz factors, $\gamma_{e,\rm min}=300$ and $\gamma_{e,\rm max}=10^5$, comparable to AGNs \citep[e.g.,][]{2013ApJ...768...54B}. Similarly, the injected electron distributions can be normalized via $\int (\gamma_em_ec^2)Q_ed\gamma_e=L_e/V$. Given the magnetic field strength $B=0.1$, we estimate the electron cooling Lorentz factor to be $\gamma_{e,c}\simeq6\pi m_ec/(t_{\rm fs}\sigma_TB^2)\sim4\times10^4$, where $\sigma_T$ is the Thomson cross section. The cooling Lorentz factor is significantly larger than $\gamma_{e,\rm min}$,  which indicates that the electrons are in the slow-cooling regime. The presence of slow-cooling electrons together with a relatively weak magnetic field would result in an enhanced inverse Compton component compared to the synchrotron emission.

In the M-IR case of AT2019dsg, the EM emissions from leptonic injections and EM cascades are illustrated as the dashed and black solid curves in Fig. \ref{fig:leptonics} (see captions for the meaning of each curve). We find that the leptonic injection predicts a more broad spectrum from radio to $\gamma$-ray bands. Moreover, the leptonic contribution (especially the inverse Compton component) is comparable to the EM cascade for a moderate leptonic loading factor $K_{e/p} \simeq 0.01$. The less efficient leptonic loading ($K_{e/p}\lesssim 0.01$) implies that the EM cascade initiated by hadronic processes could dominate the EM emissions\footnote{This conclusion is consistent with the secondary emissions in merging galaxies \citep{2019ApJ...878...76Y}.}. Meanwhile, the contributions from primary electron/positron injections could become increasingly important for efficient leptonic loading with $K_{e/p}\gtrsim 0.01$. If we take the \emph{Fermi} upper limits into account, the leptonic loading factor is constrained to be $K_{e/p}\lesssim0.01$, which falls within the expected ranges from particle acceleration simulations or the AGN models. Near this upper limit, relying solely on leptonic models or EM cascade models with purely hadronic origins is insufficient to comprehend the multi-messenger emissions from TDEs, emphasizing the necessity of a lepto-hadronic model.  One caveat in this discussion is that the limits on $K_{e/p}$ depend on the assumptions of the electron/positron distributions. If a higher $\gamma_{e,\rm min}$ or a stronger magnetic field is used, the electrons may be in the fast-cooling regime, and the synchrotron component would become increasingly important, leading to more stringent X-ray constraints. A more quantitative discussion of leptonic injections is beyond the scope of this work and will be provided in future lepto-hadronic models of TDEs.

\section{Summary and conclusions}\label{sec:summary}

In this work, we have presented a comprehensive examination of the quasi-isotropic EM cascade emissions in TDE radiation zones in a fully time-dependent manner and investigated the interconnections between the neutrino and EM cascade counterparts. We have demonstrated that in the observer's frame, EM cascades can be measured in the X-ray to $\gamma$-ray ranges at the OUV peak time or exhibit a time delay depending on the $p\gamma$ optical thickness. {For instance, if the $p\gamma$ system is optically thin and the luminosity of target photons reaches its maximum at a later time, the peak of EM cascade emission appears after the OUV peak. The resulting time delay is jointly determined by the timescale of $p\gamma$ interactions and the peak time of target photons, spanning from tens to hundreds of days. In cases where the target photon evolution with time is negligible, such as the M-IR scenario of AT2019dsg, the time delay can be solely attributed to the $p\gamma$ interaction time.}

For AT2019dsg, an X-ray data point measured around 100 days after the OUV peak, which is potentially incompatible with the early exponential decay of the X-ray flux, can be described by the X-ray cascade emission in the M-IR scenario, suggesting that the EM cascade emissions could be detectable for Swift-XRT, XMM-Newton, and NICER. Additionally, we find that if no $\gamma$-ray from the TDEs is seen, the source parameters, such as the radii of the radiation zones and the maximum energies of injected protons, can be stringently constrained. Specifically, for AT2019dsg and AT2019fdr, the \emph{Fermi} $\gamma$-ray upper limits imply that the source may not be a very efficient neutrino emitter, and the radiation zone is likely $p\gamma$ optically thin, where the predicted neutrino event rate (i.e., 0.01-0.1 neutrino events per TDE) is consistent with the expected neutrino numbers from GFU searches \citep{2021NatAs...5..510S,2022PhRvL.128v1101R}. 

The joint analysis of the EM cascades and neutrino counterparts presented in this paper provides an intriguing template for the multi-messenger studies of TDEs. Future observations of the SEDs and light curves of X-ray, $\gamma$-ray, and neutrino emissions will allow us to infer the physical conditions of the radiation zones and test our EM cascade models. In addition to the TDEs considered in this paper, the quasi-isotropic EM cascade models can be widely applied to more TDEs that do not exhibit a direct connection to relativistic jets. Moreover, EM cascades could play a significant role in jetted TDEs as well, thereby highlighting the utility of coherent lepto-hadronic modeling. Detailed discussions on this topic will be presented in our forthcoming works.

%%%%%%%%%%%%%%%%%%%%%%%%%%%%%%%%%%%%%%%%%%%%%%%%%%
%%%%%%%%%%%%%%%%%%%%%%%%%%%%%%%%%%%%%%%%%%%%%%%%%%
\acknowledgments
We would like to thank Maria Petropoulou for useful comments during the Lepto-Hadronic Workshop in Bochum, Germany, and Claire Guepin for questions motivating this work. We also extend our appreciation to Steffen Hallmann and Xin-Yue Shi for their thorough internal review.

\appendix
\section{{Interaction rates for secondary particles}}\label{app:rates}

\begin{figure}
    \centering
    \includegraphics[width=0.6\textwidth]{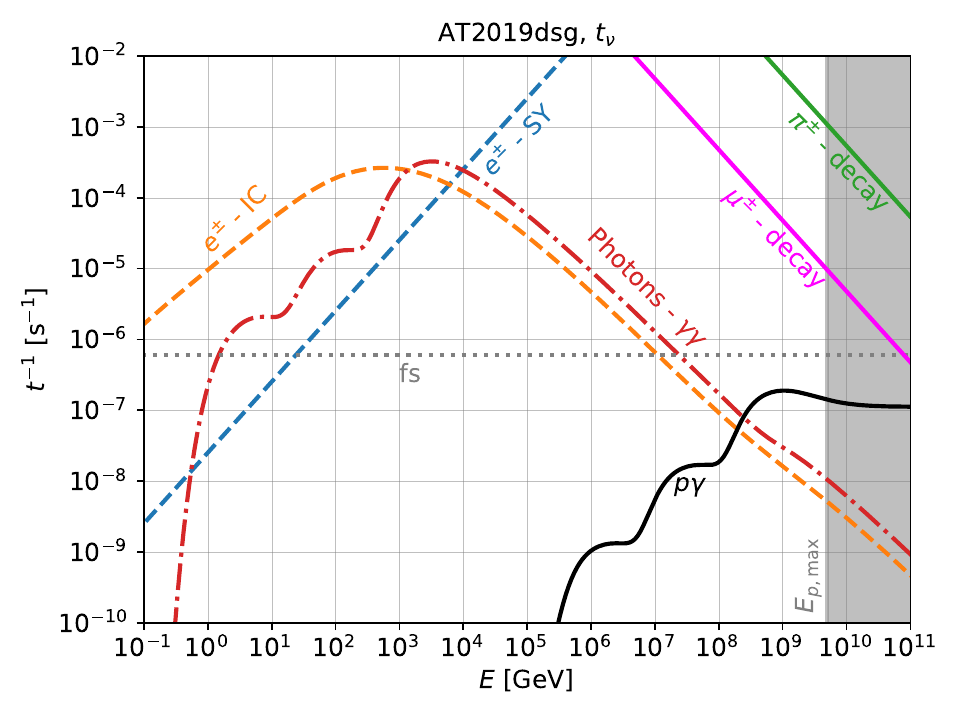}
    \caption{{Interaction rates for secondary processes, including muon decay (magenta solid), charged pion decay (green solid), $\gamma\gamma$ attenuation (red dash-dotted), electron/positron synchrotron (blue dashed) and inverse Compton (orange dashed) at the time of the neutrino emission. The $p\gamma$ and free-streaming rates are shown as black solid curve and dotted horizontal line respectively.}}
    \label{fig:cas_rates}
\end{figure}

{In Fig. \ref{fig:cas_rates}, we present the interaction rates of secondary particles contributing to the cascade emission, and compare them to the rate of the $p\gamma$ processes (depicted by the black solid curve) that initiate the electromagnetic (EM) cascade. These curves are derived within the M-IR scenario of AT2019dsg, at the neutrino detection time $t_\nu$. The decay rates of muons, for instance, $t_{\mu, \rm dec}^{-1}=(\gamma_\mu t_{\mu})^{-1}$, and charged pions, such as $t_{\pi^{\pm}, \rm dec}^{-1}=(\gamma_\pi t_{\pi^{\pm}})^{-1}$, are represented as magenta and green lines, respectively. Here, $\gamma_\mu \equiv E_\mu/(m_\mu c^2)$ stands for the Lorentz factor of muons with energy $E_\mu$, $t_{\mu}$ is the muon rest-frame lifetime, and corresponding quantities are defined for charged pions. The decay of neutral pions is exceedingly rapid and can be treated as an instantaneous process. We illustrate the synchrotron and inverse Compton rates for secondary electrons/positrons originating from $\gamma\gamma$ annihilation and muon decays using blue and orange dashed curves. Additionally, the red dash-dotted curve represents the $\gamma\gamma$ attenuation rate.}

{In general, the radiation rate for cascade emissions can be estimated as the summation of all interaction rates that build up EM cascade. From Fig. \ref{fig:cas_rates}, we noticed that the decay rates and interaction rates of all secondary particles are much larger than $p\gamma$ rate. Especially, the $\gamma\gamma$ and $p\gamma$ rates at their respective threshold energies can be connected via \citep[e.g.,][]{2016PhRvL.116g1101M,2022ApJ...932...80Y}
\begin{equation}
\frac{t_{\gamma\gamma}^{-1}}{t_{p\gamma}^{-1}}\approx\frac{\sigma_{T}}{\sigma_{p\gamma}}\frac{t_{\rm fs}^{-1}}{t_{p,\rm esc}^{-1}}\sim10^3\frac{t_{\rm fs}^{-1}}{t_{p,\rm esc}^{-1}}\gtrsim 10^3,
\end{equation}
where $\sigma_T\simeq6.65\times10^{-25}~\rm cm$ is the Thomson cross section, $\sigma_{p\gamma}\sim 5\times10^{-28}~\rm cm$ is the $p\gamma$ cross section at $\Delta$ resonance, and we applied the relation $t_{\rm fs}^{-1}/t_{p,\rm esc}^{-1}\ge1$. This analytical estimation is consistent with the curves in Fig. \ref{fig:cas_rates}. The much faster secondary processes imply that we can use the $p\gamma$ time scale to approximate the time scale to develop EM cascade. This conclusion is valid for both AT2019dsg and AT2019fdr in M-IR and M-OUV scenarios.}

\end{CJK*}

\bibliographystyle{aasjournal}
\bibliography{ref.bib}

\end{document}